\documentclass[11pt,a4paper]{article}
\usepackage{jheppub}
\usepackage{color}
\usepackage[dvipsnames]{xcolor}
\usepackage{graphicx}
\usepackage{wrapfig,enumerate,slashed}
\usepackage{subcaption}
\usepackage{multirow,enumitem}
\usepackage{footmisc}
\usepackage{amsmath}
\usepackage{graphicx}
\usepackage{hyperref}
\usepackage{enumitem}
\usepackage{tikz}
\usepackage{tikz-feynman}
\usepackage{orcidlink}
\usepackage{caption}
\usepackage{pdflscape}

\newcommand{\be}{\begin{eqnarray}}
	\newcommand{\ee}{\end{eqnarray}}

\newcommand{\bee}{\begin{eqnarray}}
	\newcommand{\eee}{\end{eqnarray}}
\newcommand{\beeq}{\begin{equation}}
	\newcommand{\eeeq}{\end{equation}}

\newcommand{\lag}{\mathcal{L}}
\newcommand{\uvlag}{\mathcal{L}_{\text{UV}}}

\newcommand{\Tr}{\text{Tr}}

\newenvironment{cedescription}{
    
	\begin{description}[leftmargin=1.2cm, style=sameline]
	}{
	\end{description}
}

\begin{document}
\flushbottom
\allowdisplaybreaks
	
%%%%%%%%%%%%%%%%%%%%%%%%%%%%%%%%%%%%%%%%%%%%%%%%%%%%%
\title{On the Robustness of type-II Seesaw Collider Searches}
%%%%%%%%%%%%%%%%%%%%%%%%%%%%%%%%%%%%%%%%%%%%%%%%%%%%%	
\author[a]{Christoph Englert\orcidlink{0000-0003-2201-0667},}
\author[c,d]{Manimala Mitra\orcidlink{0000-0002-8032-5125},}
\author[b]{Wrishik Naskar\orcidlink{0000-0002-4357-8991},}
\author[c,d]{Subham Saha\orcidlink{0009-0009-1183-3271}}
%%%%%%%%%%%%%%%%%%%%%%%%%%%%%%%%%%%%%%%%%%%%%%%%%%%%%	
\affiliation[a]{Department of Physics \& Astronomy, University of Manchester, Oxford Road,\\Manchester M13 9PL, United Kingdom}
\affiliation[b]{Deutsches Elektronen-Synchrotron DESY, Notkestra\ss{}e 85, 22607 Hamburg, Germany}
\affiliation[c]{Institute of Physics, Sachivalaya Marg, Bhubaneswar 751005, India}
\affiliation[d]{Homi Bhabha National Institute, BARC Training School Complex,\\Anushakti Nagar, Mumbai 400094, India}
%%%%%%%%%%%%%%%%%%%%%%%%%%%%%%%%%%%%%%%%%%%%%%%%%%%%%	
\emailAdd{christoph.englert@manchester.ac.uk, manimala@iopb.res.in, wrishik.naskar@desy.de, subham.saha@iopb.res.in}

%%%%%%%%%%%%%%%%%%%%%%%%%%%%%%%%%%%%%%%%%%%%%%%%%%%%%
\preprint{DESY-26-033, IOP/BBSR/2026-03}	
%%%%%%%%%%%%%%%%%%%%%%%%%%%%%%%%%%%%%%%%%%%%%%%%%%%%%	
\abstract{Electroweak triplet Higgs sector extensions are well-motivated scenarios to address lepton flavour observations. These models can also be strongly constrained by combining precise, indirect low-energy measurements with direct searches for exotic, doubly charged Higgs bosons. Together, these searches set competitive constraints on the type-II seesaw mechanism. In this work, we consider extensions of the type-II seesaw, specifically through the lens of a modified collider phenomenology. Surveying motivated extensions, we map out changes in expected correlations, focusing on the modified production and decay phenomenology of exotic Higgs particles. This enables us to assess the robustness of the type-II seesaw collider constraints against extended new-physics contributions that modify standard sensitivity expectations and projections.}
%%%%%%%%%%%%%%%%%%%%%%%%%%%%%%%%%%%%%%%%%%%%%%%%%%%%%
\maketitle

\section{Introduction}
\label{sec:intro}
%%%%%%%%%%%%%%%%%%%%%%%%%%%%%%%%%%%%%%%%%%%%%%%%%%%%%
The observation of neutrino oscillations~\cite{Kajita:2016cak,K2K:2002icj,KamLAND:2002uet,McDonald:2016ixn} offers some of the most compelling evidence for physics beyond the Standard Model (SM). Introducing a right-handed sterile neutrino allows for a direct analogue of the SM mechanism for fermion mass generation, but it also exacerbates the challenges associated with fermion flavour hierarchies. Dynamical explanations for the smallness of neutrino masses often invoke seesaw mechanisms (for a review, see~\cite{King:2003jb}). Among these, the type-II seesaw~\cite{Konetschny:1977bn,Mohapatra:1980yp,Lazarides:1980nt,Schechter:1980gr,Magg:1980ut,Cheng:1980qt} has a rich collider-relevant phenomenology. This model extends the SM with an $SU(2)_L$ triplet scalar $\Delta$ with hypercharge $Y=1$, inducing Majorana masses for neutrinos via Yukawa interactions of $\Delta$ with the SM left-handed lepton doublet ($L_L$),
\begin{equation*}
\lag \supset -{Y_{ij} } \overline{L}^{c}_{L_i}\, \Delta \, L_{L_j}\,,
\end{equation*}
with $i,j$ labelling the fermion generations. The model's Higgs-sector extension, together with lepton-number-violating interactions, makes the type-II seesaw a promising target for experimental tests across a wide range of energy scales and experimental domains. Ongoing progress in both low-energy precision experiments and high-energy collider searches continues to tighten constraints on its parameter space, e.g.~\cite{Ducu:2024xxf,Bolton:2024thn}. 

This naturally raises the question of whether the minimal type-II scenario is realised in nature with observable consequences. If embedded in a broader ultraviolet (UV) framework, correlations between observables at the LHC and low-energy experiments could deviate significantly from those predicted in the minimal model. Treating such UV completions agnostically, effective field theory (EFT) techniques can be used to explore generic modifications to the phenomenological patterns~\cite{Banerjee:2020jun,DasBakshi:2021xbl}, analogous to the Standard Model Effective Field Theory (SMEFT) approach, as demonstrated in Refs.~\cite{Barger:2008jx,Cho:2021itv,Chen:2019ebq,Cho:2023oad,Oikonomou:2024jms,Anisha:2019nzx,Banerjee:2024jwn,Crivellin:2016ihg,Birch-Sykes:2020btk,Anisha:2022hgv,Anisha:2023vvu,Ouazghour:2023plc,Anisha:2021jlz,Ashanujjaman:2022tdn,Ashanujjaman:2021txz,Padhan:2022hhl,Das:2023zby}. Furthermore, such `deformations' of the standard seesaw can also have phenomenological implications that reduce the sensitivity of experimental searches, e.g., by changing the expected final-state and its kinematics. Put another way, if a significant departure from the standard seesaw is indeed realised, current analysis strategies might miss direct hints of new physics, e.g., at the LHC. 

It is the purpose of this paper to re-analyse existing search strategies for tell-tale signs of the type-II model by means of its EFT extension. Rather than performing a global analysis (see, e.g.,~\cite{Ellis:2020unq,Giani:2023gfq,Celada:2024mcf,Durieux:2018tev,DeBlas:2019qco}), we identify key interactions that affect the dominant searches for doubly charged scalars at the LHC, acting as a phenomenological `straw man' to modify the phenomenology away from the standard type-II expectation. By tensioning the `vanilla' type-II seesaw against this phenomenologically motivated EFT generalisation, we can assess the experimental scope for sensitivity degradation. 

Of course, any EFT itself is not a meaningful theory in the sense that it directly points to a more profound theory of nature, but rather it forms a translation tool between observations and candidate theories~\cite{deBlas:2017xtg,Banerjee:2020jun, DasBakshi:2021xbl,Naskar:2022rpg,Weinberg:1978kz,Grzadkowski:2010es,Brivio:2017vri,Aebischer:2025qhh,Dawson:2021xei,Dawson:2022cmu,Ellis:2023zim,Dawson:2024ozw,Adhikary:2025gdh,Henning:2014wua, Drozd:2015rsp, Ellis:2016enq, delAguila:2016zcb, Ellis:2017jns, Kramer:2019fwz, 
Banerjee:2023iiv, Banerjee:2023xak, Chakrabortty:2023yke, Cohen:2020fcu, Dittmaier:2021fls}. Therefore, to clarify the experimental robustness of type-II seesaw analysis strategies, we provide a survey of potential UV extensions that dynamically introduce the interactions we consider primarily for their phenomenological merits. In turn, this creates a juxtaposition of theoretical motivation and experimental sensitivity.

This work is organised as follows. In Sec.~\ref{sec:EFTops}, we discuss the theoretical motivation of our analysis. We first briefly review the type-II seesaw model to make this work self-consistent. More attention is given in Sec.~\ref{sec:drive} to motivating the specific EFT interactions considered in this work. These modify the collider phenomenology of type-II scalar exotics, and are particularly relevant for tensioning the model against the most sensitive direct collider searches. Section~\ref{sec:uvcomp} clarifies how these operators can arise from concrete particle extensions of the type-II seesaw. Section~\ref{sec:robust} is devoted to obtaining a quantitative understanding of the robustness of current type-II search strategies: Section~\ref{sec:pheno} discusses the phenomenological implications of Sec.~\ref{sec:EFTops}, Sec.~\ref{sec:impl} clarifies their implication for existing exclusion constraints, and Sec.~\ref{sec:discov} details the relevance for the discovery of type-II states in the future. We summarise and conclude in Sec.~\ref{sec:conc}.

%%%%%%%%%%%%%%%%%%%%%%%%%%%%%%%%%%%%%%%%%%%%%%%%%
\section{Deforming the Type-II Seesaw Mechanism}
\label{sec:EFTops}
%%%%%%%%%%%%%%%%%%%%%%%%%%%%%%%%%%%%%%%%%%%%%%%%%
%%%%%%%%%%%%%%%%%%%%%%%%%%%%%%%%%%%%%%%%%%%%%%%%%
\subsection{LHC Phenomenology and Relevant Effective Interactions}
\label{sec:drive}
%%%%%%%%%%%%%%%%%%%%%%%%%%%%%%%%%%%%%%%%%%%%%%%%%
The type-II seesaw mechanism is a simple and well-motivated extension of the SM that provides a natural explanation for the origin of Majorana neutrino masses. In this framework, lepton number violation arises through the introduction of an additional scalar field $\Delta$ that transforms as $\Delta \sim (\mathbf{1},\mathbf{3},1)$ under the SM gauge group $SU(3)_C \times SU(2)_L \times U(1)_Y$. $\Delta$ can be conveniently expressed in matrix form as
\begin{equation}
\Delta
= \frac{1}{\sqrt{2}}
\begin{pmatrix}
\delta^+ & \sqrt{2}\Delta^{++} \\
v_\Delta + \delta^0 + i \xi^0 & -\delta^+
\end{pmatrix}\,,
\end{equation}
where $v_\Delta$ denotes the vacuum expectation value (vev) of the neutral component, and $\delta^0$ and $\xi^0$ represent the CP-even and CP-odd neutral fluctuations, respectively. The kinetic term of the triplet field takes the standard gauge-covariant form
\begin{equation}
\lag \supset \text{Tr}\!\left[(D_\mu \Delta)^\dagger D^\mu \Delta \right].
\end{equation}
The scalar potential involving the SM Higgs doublet $\Phi$ and the scalar triplet $\Delta$ can be written as
\begin{equation}
\label{eqn:scalarpot}
    \begin{split}
        V(\Phi,\Delta) = -m_H^2 \,(\Phi^\dagger \Phi) + \lambda_H (\Phi^\dagger \Phi)^2 
        + m_\Delta^2\, \text{Tr} [\Delta^\dagger \Delta] 
        + \lambda_{\Delta_1}\, (\text{Tr} [\Delta^\dagger \Delta])^2 
        + \lambda_{\Delta_2}\, \text{Tr} [(\Delta^\dagger \Delta)^2] \\
        + \lambda_{\Delta_3}\, (\Phi^\dagger \Phi)\, \text{Tr} [\Delta^\dagger \Delta] 
        + \lambda_{\Delta_4}\, \Phi^\dagger \Delta \Delta^\dagger \Phi 
        + \left[ \mu_{\Delta}\, \Phi^\dagger i\sigma^2 \Delta^\dagger \Phi + \text{h.c.}\right]\,.
    \end{split}
\end{equation}
After electroweak symmetry breaking, both the doublet and triplet acquire vevs. Minimising the potential yields the relations
\begin{align}
m_H^2 &= \lambda_H v^2 - \sqrt{2}\, \mu_\Delta v_\Delta 
        + \frac{1}{2}(\lambda_{\Delta_3} + \lambda_{\Delta_4}) v_\Delta^2\,,\\
m_\Delta^2 &= \frac{\mu_\Delta v^2}{\sqrt{2} v_\Delta} 
             - \frac{1}{2}(\lambda_{\Delta_3} + \lambda_{\Delta_4}) v^2 
             - (\lambda_{\Delta_1} + \lambda_{\Delta_2}) v_\Delta^2\,.
\end{align}
The mass spectrum of the physical scalar states can then be obtained by expanding the potential around the vacuum configuration as usual~\cite{Banerjee:2024jwn}.
To leading order in $v_\Delta / v \ll 1$, one finds
\begin{align}
M_{\Delta^{\pm\pm}}^2 &\simeq m_\Delta^2 - \frac{\lambda_{\Delta_4}}{2} v^2\,, \\
M_{\Delta^\pm}^2 &\simeq m_\Delta^2 - \frac{\lambda_{\Delta_4}}{4} v^2\,, \\
M_{A}^2 &\simeq m_\Delta^2\,, \\
M_{\Delta^0,\,h}^2 &\simeq
\begin{pmatrix}
2\lambda_H v^2 & -\sqrt{2}\, \mu_\Delta v \\
-\sqrt{2}\, \mu_\Delta v & m_\Delta^2
\end{pmatrix}\,.
\end{align}
The mixing angle $\alpha$ between the neutral CP-even states, defined by
\begin{equation}
\tan 2\alpha \simeq 
\frac{2 \sqrt{2}\, \mu_\Delta v}{m_\Delta^2 - 2 \lambda_H v^2}\,,
\end{equation}
is typically small, ensuring that the lighter mass eigenstate $h$ remains SM-like, while the heavier state $\Delta^0$ is mostly triplet-like. The smallness of this mixing also ensures that deviations of the SM Higgs couplings from their observed values remain within current experimental limits~\cite{Primulando:2019evb}. The extended scalar sector therefore gives rise to several new physical states: a pair of doubly-charged scalars $\Delta^{\pm\pm}$, singly charged scalars $\Delta^{\pm}$, a CP-odd neutral pseudoscalar $A$, and two CP-even neutral scalars $\Delta^0,~h$. The former of these scalar exotics is a smoking-gun signature of the model, which has been extensively searched for at the LHC (more details are provided below).

The triplet couples to the lepton doublets via a Yukawa interaction
\begin{equation}
\label{eq:yuk2}
\lag \supset -{Y_{ij}}\, \overline{L}^{c}_{L_i}\, \Delta \, L_{L_j} + \text{h.c.}\,,
\end{equation}
where $Y_{ij}$ is a symmetric matrix in flavour space. Once the neutral component of the triplet acquires a vev, the interaction in Eq.~\eqref{eq:yuk2} generates Majorana masses for the neutrinos
\begin{equation}
(m_\nu)_{ij} = \sqrt{2}Y_{ij}v_\Delta\,.
\end{equation}

The size of $v_\Delta$ is strongly constrained by electroweak precision data~\cite{Primulando:2019evb,Antusch:2018svb}, most notably through its contribution to the $\rho$ parameter, which at tree level reads as $\rho \simeq 1 - 2 v_\Delta^2 / v^2$. The current experimental value of the $\rho$ parameter implies $v_\Delta \lesssim \mathcal{O}(1\,\text{GeV})$. This upper bound, together with the observed neutrino mass scale, constrains the magnitude of $Y_{ij}$. At the same time, the relative flavour structure of the Yukawa matrix governs the branching fractions of the doubly-charged scalar into different dilepton flavour combinations~\cite{Garayoa:2007fw,Fuks:2019clu}.

From the collider perspective, the doubly-charged scalar is the most accessible state of the triplet~\cite{FileviezPerez:2008jbu,Chakrabortty:2015zpm,Primulando:2019evb,Fuks:2019clu,Cai:2017mow,Antusch:2018svb,BhupalDev:2018tox,delAguila:2013mia,ATLAS:2017xqs,CMS:2022cbe,ATLAS:2022pbd}. As discussed in the literature, the HL-LHC can probe $\Delta^{\pm\pm}$ masses up to about $1.5$ TeV through pair production, e.g.~\cite{Mitra:2016wpr,Banerjee:2024jwn,CMS:2022cbe,Li:2018jns}. This region is consistent with constraints from electroweak precision tests and Higgs data, which still allow for relatively light triplet states~\cite{Primulando:2019evb}. However, stringent limits from charged lepton flavour violating processes~\cite{Dinh:2012bp,Chakrabortty:2012vp,Barrie:2022ake}, such as $\mu \to e\gamma$ or $\mu \to 3e$, can typically require $\Delta^{\pm\pm}$ masses above $1.6$~TeV in order to remain compatible with current bounds. This poses challenges in the detection of $\Delta^{\pm \pm}$ at the HL-LHC. 

Such a situation motivates considering possible extensions of the minimal type-II seesaw scenario. An EFT framework allows one to parameterise the impact of higher-dimensional operators that may arise from UV completions agnostically. Dimension-six operators can affect both the production and decay properties of $\Delta^{\pm\pm}$, e.g., via enhancing the pair-production cross section at hadron colliders. Furthermore, modified interactions can enhance less-abundant decay channels, reducing sensitivity in the `standard' search channels and shifting the discovery potential toward rarer, and therefore less constrained,~ones.

In the following, we consider a subset of operators from the complete basis of Ref.~\cite{Banerjee:2020jun} that are directly relevant to the key phenomenological aspects of collider searches: (i) operators that modify the pair production of charged scalars and (ii) operators that induce new decay channels of the doubly-charged scalar, such as $\Delta^{\pm\pm}\rightarrow \ell^{\pm}\ell^{\pm}\gamma$, thereby altering its decay phenomenology.

For charged-scalar pair production, the relevant operators include $\mathcal{O}_{G\Delta}$, $\mathcal{O}_{\tilde{G}\Delta}$, $\mathcal{O}_{Q \Delta \mathcal{D}}^{(1)}$, $\mathcal{O}_{Q \Delta \mathcal{D}}^{(2)}$, $\mathcal{O}_{u\Delta D}$, $\mathcal{O}_{d\Delta D}$, $\mathcal{O}_{Q d H \Delta}^{(1)}$, $\mathcal{O}_{Q u H \Delta}^{(1)}$, $\mathcal{O}_{Q d H \Delta}^{(2)}$, and $\mathcal{O}_{Q u H \Delta}^{(2)}$. Among these, we focus on $\mathcal{O}_{G\Delta}$, $\mathcal{O}_{Q \Delta \mathcal{D}}^{(1)}$, and $\mathcal{O}_{u\Delta D}$ as representative examples to illustrate the qualitative differences between gluon-initiated and quark-initiated production mechanisms. Although the analyses presented in Sec.~\ref{sec:impl} and the subsequent sections are performed considering only the $\mathcal{O}_{G\Delta}$ operator as a representative candidate. The explicit forms of these operators are given in Eqs.~\eqref{eq:OGDelta}, \eqref{eq:QDeltaD}, and \eqref{eq:uDeltaD}, respectively.

Similarly, the operators $\mathcal{O}_{LeH\Delta D}$ and $\mathcal{O}_{BL\Delta}$ contribute to the decay channel $\Delta^{\pm\pm}\rightarrow \ell^{\pm}\ell^{\pm}\gamma$. Their explicit forms are given in Eqs.~\eqref{eq:LeHDeltaD} and \eqref{eq:BLDelta}, respectively. In this work, we use $\mathcal{O}_{BL\Delta}$ as a representative operator to investigate the phenomenological implications of this decay mode.
\begin{align}
\label{eq:OGDelta}
\mathcal{O}_{G\Delta}&=G^A_{\mu \nu}G^{A\mu\nu}~\Tr[\Delta^{\dagger}\Delta]\,,\\
\label{eq:uDeltaD}
\mathcal{O}_{u\Delta D}&=N^2_f\bar{u}_{p\alpha}\gamma^{\mu}u^{\alpha}_q~\Tr[\Delta^{\dagger} i \overset{\leftrightarrow}{\mathcal{D}}_{\mu}
 \Delta ]\,,\\
\label{eq:BLDelta}
 \mathcal{O}_{BL\Delta}&=\frac{1}{2}(N^2_f-N_f)(L^T_pCi\tau_2\Delta\sigma_{\mu\nu}L_q)\,B^{\mu\nu}\,,\\
\label{eq:LeHDeltaD}
 \mathcal{O}_{LeH\Delta D}&=N^2_f~\Tr[L^T_pCi\tau_2(\gamma^{\mu}D_{\mu}\Delta) H e_q]\,,\\
\label{eq:QDeltaD}
\mathcal{O}_{Q \Delta \mathcal{D}}^{(1)}& =
	N_f^2 \,(\overline{Q}_{p\alpha i} \,\gamma^{\mu} \,Q^{\alpha i}_{q})~\Tr[(\Delta^{\dagger} \,i\overleftrightarrow{\mathcal{D}}_{\mu} \,\Delta)]\,.
\end{align}
This operator set is not exhaustive at the dimension-six level~\cite{Banerjee:2020jun}, but it will enable us to transparently contextualise theoretical motivation, experimental sensitivity expectations, and modifications. Here we follow Ref.~\cite{Banerjee:2020jun}, where the superscript $(1)$ in Eq.~\eqref{eq:QDeltaD} denotes a particular class of operators with the specified field content, among several possible choices. The index $N_f$ labels the number of fermion flavours in the theory. 

Of course, $\Delta ^{\pm \pm} \to l^{\pm} l^{\pm} \gamma$ arises as a real-emission contribution as part of the QED radiative correction in the standard seesaw. With Eq.~\eqref{eq:BLDelta} resulting from integrating out heavy states, this interaction will not modify the soft and collinear behaviour of QED corrections (see, e.g.,~\cite{Englert:2018byk}), but rather induce a redistribution of the decay phenomenology towards comparably hard photons. Typical searches for doubly-charged scalars include lepton isolation criteria and exploit the back-to-back decay structure of prompt leptonic decays of the $\Delta^{\pm\pm}$, produced at low velocity for high-mass exclusion. A significant modification of the two-body decay towards three-body decay kinematics might therefore reduce the sensitivity of these established searches, especially in the high-mass region. Also, the size of the effective operator contribution to create such an outcome then compares directly to the perturbative expansion of the standard doubly-charged Higgs scenario without dimension-six terms. This comparison can then be framed as a criterion for the perturbative robustness of the existing strategy and its expected sensitivity. A similar argument can be made for the additional decay-relevant operators that dilute the branching ratio into same-sign di-lepton final states.

Coming back to the operators that change the production probability such as $\mathcal{O}_{G\Delta}$, again, a perturbativity argument can be constructed: If the effective modification to escape the experimental sensitivity needs to be large, we force the theory towards the strong coupling limit, where we cannot trust our theoretical modelling in the first~place. 

%%%%%%%%%%%%%%%%%%%%%%%%%%%%%%%%%%%%%%%%%%%%%%%%%%%%%
\subsection{UV Completing the EFT Operators}
\label{sec:uvcomp}
%%%%%%%%%%%%%%%%%%%%%%%%%%%%%%%%%%%%%%%%%%%%%%%%%%%%%
While the impact of these effective operators can be quantified directly at the phenomenological level through their effect on production cross sections and branching ratios, it is equally important to understand how such structures could emerge from a more consistent, first-principles UV framework. Identifying possible UV completions not only provides theoretical justification for the presence of these operators, but also constrains the range of viable model-building options. First, we examine the operators that directly enhance the pair production of the doubly-charged scalars and clarify the gauge charges of potential UV completions.\footnote{We use \texttt{GroupMath}~\cite{Fonseca:2020vke} to check if gauge charges are conserved at the respective vertices.}

%%%%%%%%%%%%%%%%%%%%%%%%%%%%%%%%%%%%%%%%%%%%%%%%%%%%%
\subsection*{Operator $\mathcal{O}_{G \Delta}$}
%%%%%%%%%%%%%%%%%%%%%%%%%%%%%%%%%%%%%%%%%%%%%%%%%%%%%
Given the gluonic initial states, we first turn to colour-charged particles, such as vector-like quarks (VLQs) or non-trivially charged scalars, as possible mediators. The relevant Feynman topologies and the corresponding gauge charges of these states are outlined below.
\begin{itemize}
    \item Allowing heavy–light mixing with the SM left-handed quark doublet $Q_L~(\mathbf{3},\mathbf{2},1/6)$, one possible choice is a vector-like quark (VLQ)
    $\color{red}X_L~(\mathbf{\bar{3}},\mathbf{2},-7/6)$, which couples through terms in the UV Lagrangian of the form
    \begin{equation*}
        \uvlag \supset y^\prime_{X} (\bar{X}_L^c \Delta Q_L + \text{h.c.}) \, .
    \end{equation*}
    Alternatively, one may introduce a VLQ 
    $\color{red}X_L~(\mathbf{3},\mathbf{2},7/6)$, yielding
    \begin{equation*}
        \uvlag \supset y^\prime_{X} (\bar{X}_L \Delta Q_L + \text{h.c.}) \, .
    \end{equation*}
    Mixing with right-handed up- and down-type singlets leads to VLQs in the triplet representation of $SU(2)_L$, ${\color{red}{X_L~(\mathbf{3/\bar{3}},\mathbf{3},\mp 5/3)}}$, ${\color{red}{(\mathbf{3/\bar{3}},\mathbf{3},\mp 2/3)}}$,
    with corresponding interactions
    \begin{equation}
        \uvlag \supset y^\prime_{X} (\bar{X}^{/c}_L \Delta u_R + \text{h.c.}) 
        \quad \text{or} \quad 
        y^\prime_{X} (\bar{X}^{/c}_L \Delta d_R + \text{h.c.}) \, .
    \end{equation}
    These interactions give rise to the production mechanisms illustrated in the following topologies
    \begin{equation*}
    		\begin{tikzpicture}[baseline=(a)]
			\begin{feynman}
				\vertex (a) at (0, 0);
				\vertex (b) at (1, 0);
				\vertex (c) at (1, 1) ;
				\vertex (d) at (0, 1) ;
				\vertex (g1) at (-1, 0) {\(\large g\)};
				\vertex (g2) at (-1, 1) {\(\large g\)};
				\vertex (d1) at (2, 1) {\(\large \Delta^{++}\)};
				\vertex (d2) at (2, 0) {\(\large \Delta^{--}\)};
				
				\diagram* {
					(d) -- [fermion] (a) -- [fermion] (b),
					(d) -- [anti fermion] (c) -- [anti fermion, thick, red] (b),
					(g1) -- [gluon] (a),
					(g2) -- [gluon] (d),
					(d1) -- [scalar] (c),
					(d2) -- [scalar] (b),
				};
			\end{feynman}
		\end{tikzpicture} \quad \begin{tikzpicture}[baseline=(a)]
			\begin{feynman}
				\vertex (a) at (0, 0);
				\vertex (b) at (1, 0);
				\vertex (c) at (1, 1) ;
				\vertex (d) at (0, 1) ;
				\vertex (g1) at (-1, 0) {\(\large g\)};
				\vertex (g2) at (-1, 1) {\(\large g\)};
				\vertex (d1) at (2, 1) {\(\large \Delta^{++}\)};
				\vertex (d2) at (2, 0) {\(\large \Delta^{--}\)};
				
				\diagram* {
					(d) -- [fermion, thick, red] (a) -- [fermion, thick, red] (b),
					(d) -- [anti fermion, thick, red] (c) -- [anti fermion] (b),
					(g1) -- [gluon] (a),
					(g2) -- [gluon] (d),
					(d1) -- [scalar] (c),
					(d2) -- [scalar] (b),
				};
			\end{feynman}
		\end{tikzpicture}\,~.
        \end{equation*}
	\item Scalars carrying non-trivial $SU(3)_C$ charges, i.e. fields 
    $\color{red}S~(\mathbf{R}_C,\mathbf{R}_L,Y)$ with $\mathbf{R}_C \neq \mathbf{1}$, can also generate $\mathcal{O}_{G \Delta}$ when integrated out. In this case, the UV Lagrangian contains operators of the form 
\begin{equation}
        \uvlag \supset\text{Tr}(S^\dagger S) \, \text{Tr}(\Delta^\dagger \Delta) \, .
\end{equation}
The resulting topology contributing to the pair production of $\Delta^{\pm \pm}$ is   
\begin{equation*}
		\begin{tikzpicture}
			\begin{feynman}
				\vertex (a) at (0, 0);
				\vertex (b) at (0.87, 0.5);
				\vertex (d) at (0, 1) ;
				\vertex (g1) at (-1, 0) {\(\large g\)};
				\vertex (g2) at (-1, 1) {\(\large g\)};
				\vertex (d1) at (1.73, 1) {\(\large \Delta^{++}\)};
				\vertex (d2) at (1.73, 0) {\(\large \Delta^{--}\)};
				
				\diagram* {
					(d) -- [scalar, thick, red] (a) -- [scalar, thick, red] (b),
					(d) -- [scalar, thick, red] (b),
					(g1) -- [gluon] (a),
					(g2) -- [gluon] (d),
					(d1) -- [scalar] (b),
					(d2) -- [scalar] (b),
				};
			\end{feynman}
		\end{tikzpicture}\,~.
\end{equation*}
\end{itemize}

These scenarios, broadly speaking, are compatible with frameworks based on partial compositeness, in which the SM fermions mix linearly with composite-sector operators and coloured vector-like fermionic partners are generically expected~\cite{Kaplan:1991dc,Contino:2003ve,Agashe:2004rs,Contino:2010rs,Panico:2015jxa}. Coloured scalar resonances are, by contrast, not a generic prediction of minimal partial-compositeness constructions. They can appear in concrete strongly-coupled completions (for example, as coloured pNGBs or higher resonances in models with extended cosets or non-minimal underlying dynamics~\cite{Redi:2012ha,Vecchi:2015fma}). Finally, in GUT embeddings, the scalar triplet of the type-II seesaw and additional colour-charged scalars can stem from the same larger multiplet, so a single UV multiplet can, in principle, contain both the seesaw triplet and extra coloured states~\cite{Georgi:1974sy,Fritzsch:1974nn,Bajc:2007zf}.

%%%%%%%%%%%%%%%%%%%%%%%%%%%%%%%%%%%%%%%%%%%%%%%%%%%%%
\subsection*{Operators $\mathcal{O}_{Q \Delta D}$ and $\mathcal{O}_{u \Delta D}$}
%%%%%%%%%%%%%%%%%%%%%%%%%%%%%%%%%%%%%%%%%%%%%%%%%%%%%
A simple UV completion for both of these operators involves a heavy neutral vector boson, for instance, a $\color{red}Z^\prime$, mediating the production channel
\begin{equation*}
	\begin{tikzpicture}[baseline=(a)]
        \begin{feynman}
            \vertex (a) at (0, 0);
            \vertex (b) at (1, 0);
            \vertex (q1) at (-0.7, 0.7) {\(\large q\)};;
            \vertex (q2) at (-0.7, -0.7) {\(\large \bar{q}\)};;
            \vertex (d1) at (1.7, 0.7) {\(\large \Delta^{++}\)};
            \vertex (d2) at (1.7, -0.7) {\(\large \Delta^{--}\)};
            
            \diagram* {
                (a) -- [boson, thick, red, edge label=$Z^\prime$] (b),
                (q1) -- [fermion] (a) -- [fermion] (q2),
                (d1) -- [scalar] (b) -- [scalar] (d2),
            };
        \end{feynman}
    \end{tikzpicture}\,~.
\end{equation*}
Such states arise naturally in extensions of the SM with an additional $U(1)^\prime$ gauge symmetry~\cite{Langacker:2008yv} (which have recently attracted increased interest in relation to flavour anomalies~\cite{Bonilla:2017lsq,Davighi:2021oel,Allanach:2023uxz,Allanach:2024ozu,Allanach:2020kss,Dawson:2024ozw}). In these constructions, the $Z^\prime$ couples to the SM quarks and the scalar triplet, allowing the operators $\mathcal{O}_{Q \Delta D}$ and $\mathcal{O}_{u \Delta D}$ to be generated upon integrating out the heavy vector boson. 

%%%%%%%%%%%%%%%%%%%%%%%%%%%%%%%%%%%%%%%%%%%%%%%%%%%%%
\subsection*{The decay $\Delta^{\pm \pm} \to \ell^\pm \ell^\pm \gamma$}
%%%%%%%%%%%%%%%%%%%%%%%%%%%%%%%%%%%%%%%%%%%%%%%%%%%%%
This decay process already proceeds via tree-level topologies in the minimal type-II seesaw model. The coloured scalars and VLQs introduced above contribute only at two-loop order or higher and are therefore not expected to play a significant role. Heavy neutral vector bosons (for example, a $Z^\prime$), as well as scalar extensions trivially charged under $SU(3)_C$, with doubly-charged scalars, however, can generate additional contributions, as shown in the following diagrams
\begin{equation*}
	\begin{tikzpicture}[baseline=(a)]
		\begin{feynman}
			\vertex (a) at (1, 0) {\(\large \Delta^{\pm \pm}\)};
			\vertex (b) at (2, 0);
			\vertex (t1) at (3, 0.7);
			\vertex (t3) at (2.5, -0.35);
			\vertex (g4) at (2.5, -1.35) {\(\large \gamma\)};
			\vertex (t2) at (3, -0.7);
			\vertex (w1) at (4, 0.7) {\(\large \ell^{\pm}\)};
			\vertex (b1) at (4, -0.7) {\(\large \ell^{\pm}\)};
			
			\diagram* {
				(a) -- [scalar] (b),
				(b) -- [fermion] (t1),
				(b) -- [fermion] (t2),
				(t1) -- [boson,red,thick] (t2),
				(t1) -- [fermion] (w1),
				(t2) -- [fermion] (b1),
				(t3) -- [boson] (g4),
			};
		\end{feynman}
	\end{tikzpicture}\qquad\begin{tikzpicture}[baseline=(a)]
		\begin{feynman}
			\vertex (a) at (1, 0) {\(\large \Delta^{\pm \pm}\)};
			\vertex (b) at (2, 0);
			\vertex (t1) at (3, 0.7);
			\vertex (t3) at (2.5, -0.35);
			\vertex (g4) at (2.5, -1.35) {\(\large \gamma\)};
			\vertex (t2) at (3, -0.7);
			\vertex (w1) at (4, 0.7) {\(\large \ell^{\pm}\)};
			\vertex (b1) at (4, -0.7) {\(\large \ell^{\pm}\)};
			
			\diagram* {
				(a) -- [scalar] (b),
				(b) -- [boson, edge label=$Z/\gamma$] (t1),
				(b) -- [scalar,red,thick] (t2),
				(t2) -- [fermion] (t1),
				(t1) -- [fermion] (w1),
				(t2) -- [fermion] (b1),
				(t3) -- [boson] (g4),
			};
		\end{feynman}
	\end{tikzpicture}\,~.
\end{equation*}

%%%%%%%%%%%%%%%%%%%%%%%%%%%%%%%%%%%%%%%%
\section{Collider Robustness}
\label{sec:robust}
%%%%%%%%%%%%%%%%%%%%%%%%%%%%%%%%%%%%%%%%
\subsection{Phenomenological Impact}
\label{sec:pheno}
\begin{figure}[t!]
    \centering
    % ===================== (a) =====================
    \begin{minipage}[t]{0.24\textwidth}
        \centering
        \textbf{type-II}\\[0.2cm]

        \begin{tikzpicture}[baseline={(0,0)}]
            \begin{feynman}
                \vertex (a) at (0,0);
                \vertex (b) at (-1.0, 1.0);
                \vertex (c) at (-1.0, -1.0);
                \vertex (f) at (1.5, 0);
                \vertex (d) at (2.5, 1.0);
                \vertex (e) at (2.5, -1.0);
                \diagram*{
                    (b) -- [fermion, edge label=$\bar{q}$] (a) -- [fermion, edge label'=$q$] (c),
                    (a) -- [boson, edge label=$Z/\gamma^*$] (f),
                    (f) -- [scalar, edge label'=$\Delta^{++}$] (d),
                    (f) -- [scalar, edge label=$\Delta^{--}$] (e),
                };
            \end{feynman}
        \end{tikzpicture}
    \end{minipage}
    \hfill
    % ===================== (b) =====================
    \begin{minipage}[t]{0.24\textwidth}
        \centering
        \textbf{$\mathcal{O}_{G\Delta}$}\\[0.2cm]
        \begin{tikzpicture}[baseline={(0,0)}]
            \begin{feynman}
                \vertex[blob, draw=red, fill=red, minimum size=3mm] (a) at (0,0) {};
                \vertex (b) at (-1.0, 1.0);
                \vertex (c) at (-1.0, -1.0);
                \vertex (d) at (1.0, 1.0);
                \vertex (e) at (1.0, -1.0);
                \diagram*{
                    (b) -- [gluon, edge label=$g$] (a) -- [gluon, edge label'=$g$] (c),
                    (a) -- [scalar, edge label'=$\Delta^{++}$] (d),
                    (a) -- [scalar, edge label=$\Delta^{--}$] (e),
                };
            \end{feynman}
        \end{tikzpicture}
    \end{minipage}
    \hfill
    % ===================== (c) =====================
    \begin{minipage}[t]{0.24\textwidth}
        \centering
        \textbf{$\mathcal{O}_{Q\Delta\mathcal{D}}^{(1)}$}\\[0.2cm]
        \begin{tikzpicture}[baseline={(0,0)}]
            \begin{feynman}
                \vertex[blob, draw=green, fill=green, minimum size=3mm] (a) at (0,0) {};
                \vertex (b) at (-1.0, 1.0);
                \vertex (c) at (-1.0, -1.0);
                \vertex (d) at (1.0, 1.0);
                \vertex (e) at (1.0, -1.0);
                \diagram*{
                    (b) -- [fermion, edge label=$\bar{q}$] (a) -- [fermion, edge label'=$q$] (c),
                    (a) -- [scalar, edge label'=$\Delta^{++}$] (d),
                    (a) -- [scalar, edge label=$\Delta^{--}$] (e),
                };
            \end{feynman}
        \end{tikzpicture}
    \end{minipage}
    \hfill
    % ===================== (d) =====================
    \begin{minipage}[t]{0.24\textwidth}
        \centering
        \textbf{$\mathcal{O}_{u\Delta\mathcal{D}}$}\\[0.2cm]
        \begin{tikzpicture}[baseline={(0,0)}]
            \begin{feynman}
                \vertex[blob, draw=cyan, fill=cyan, minimum size=3mm] (a) at (0,0) {};
                \vertex (b) at (-1.0, 1.0);
                \vertex (c) at (-1.0, -1.0);
                \vertex (d) at (1.0, 1.0);
                \vertex (e) at (1.0, -1.0);
                \diagram*{
                    (b) -- [fermion, edge label=$\bar{q}_u$] (a) -- [fermion, edge label'=$q_u$] (c),
                    (a) -- [scalar, edge label'=$\Delta^{++}$] (d),
                    (a) -- [scalar, edge label=$\Delta^{--}$] (e),
                };
            \end{feynman}
        \end{tikzpicture}
    \end{minipage}
    \captionsetup{justification=raggedright,singlelinecheck=false}
    \caption{Representative Feynman diagrams illustrating the pair production of $\Delta^{\pm\pm}$ in vanilla type-II seesaw and in presence of different EFT operators. The left-most panel corresponds to the vanilla type-II seesaw, while the colored blobs in other panels indicate contributions from $\mathcal{O}_{G\Delta}$ (red), $\mathcal{O}_{Q\Delta\mathcal{D}}^{(1)}$ (green), and $\mathcal{O}_{u\Delta\mathcal{D}}^{(1)}$ (cyan).}
    \label{Fig:Delta_production_feyn}
\end{figure}
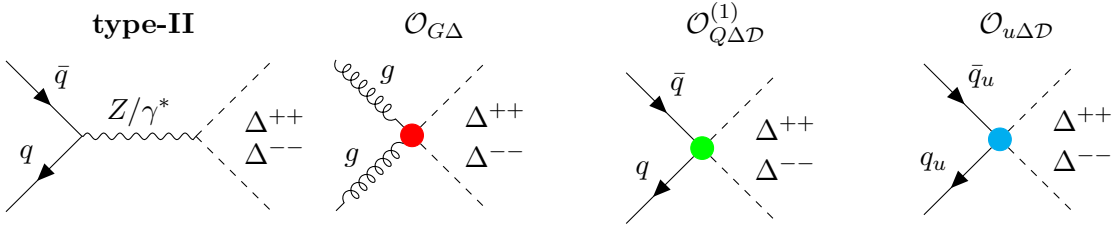

%%%%%%%%%%%%%%%%%%%%%%%%%%%%%%%%%%%%%%%%
Among the EFT operators introduced in Sec.~\ref{sec:EFTops}, $\mathcal{O}_{G\Delta}$, $\mathcal{O}_{Q \Delta \mathcal{D}}^{(1)}$ and $\mathcal{O}_{u\Delta D}$ can directly influence the pair production of doubly-charged Higgs bosons in $pp$ collisions; representative Feynman diagrams are shown in Fig.~\ref{Fig:Delta_production_feyn}. From left to right, these label contributions to the $pp \rightarrow \Delta^{++}\Delta^{--}$ production processes arising from the vanilla type-II (electroweak) mechanism, the $\mathcal{O}_{G\Delta}$ operator, the $\mathcal{O}_{Q\Delta\mathcal{D}}^{(1)}$ operator, and the $\mathcal{O}_{u\Delta\mathcal{D}}^{(1)}$ operator, respectively. Here, $q_{u}$ denotes an up-type quark.  

\begin{figure}[!t]
    \centering
    % First figure
    \begin{subfigure}{0.49\textwidth}
        \centering
       
        \includegraphics[width=\textwidth]{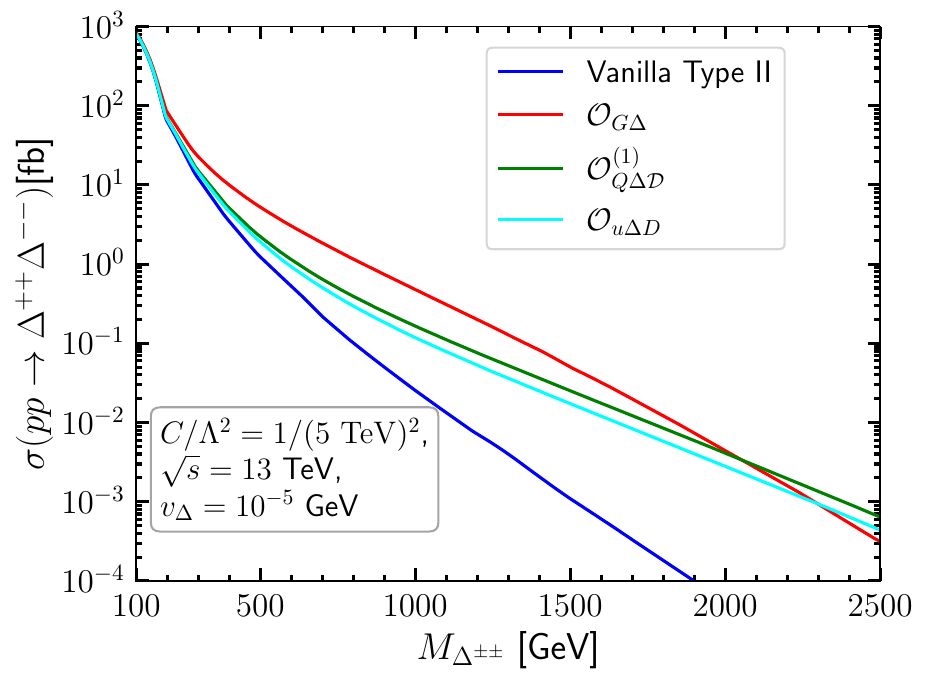}
        \caption{}
        \label{fig:Prod_HPPHMM}
    \end{subfigure}
    % Second figure
    \begin{subfigure}{0.49\textwidth}
        \centering
       
        \includegraphics[width=\textwidth]{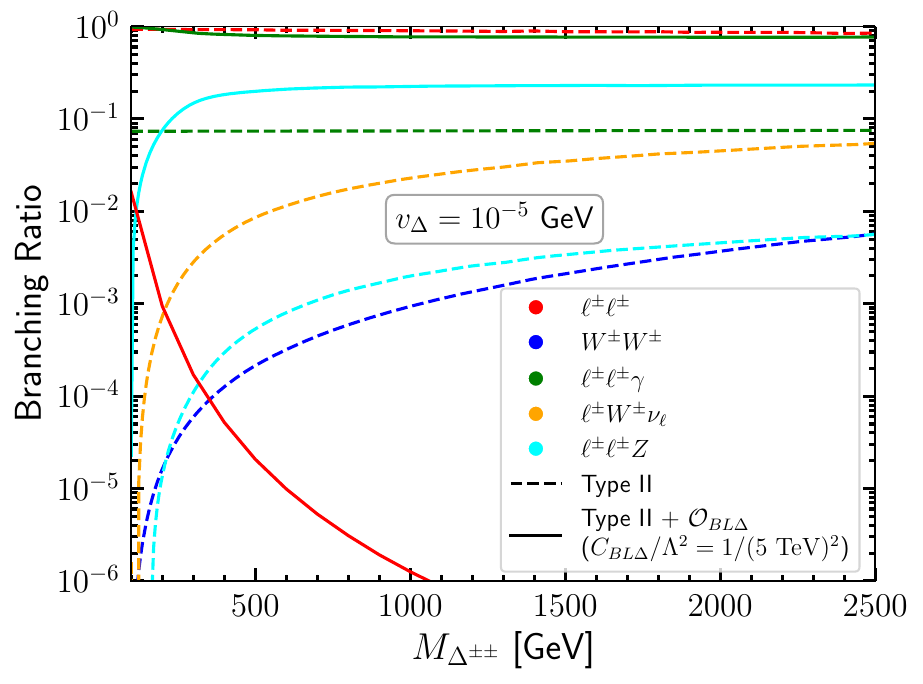}
        \caption{}
        \label{fig:BR_Delta}
    \end{subfigure}
    \caption{(a) Production cross section of $pp \rightarrow \Delta^{++} \Delta^{--}$ for vanilla type-II and different EFT operator contributions. (b) Branching ratios of different decay modes of $\Delta^{\pm \pm}$ as a function of $M_{\Delta^{\pm \pm}}$. }
    \label{fig:Mll_dis}
\end{figure}

To compute cross sections, we implement the type-II seesaw model and its EFT extension (including the operators given in Sec.~\ref{sec:EFTops}) on \texttt{FeynRules}~\cite{Alloul:2013bka}, which is interfaced with \texttt{MadGraph5\_aMC@NLO-v3.5.3}~\cite{Alwall:2014hca} through a \texttt{Ufo}~\cite{Degrande:2011ua} model. In Fig.~\ref{fig:Prod_HPPHMM}, we present the variation of the production cross section of $pp \rightarrow \Delta^{++}\Delta^{--}$ with $M_{\Delta^{\pm\pm}}$ at a $\sqrt{s}=13$~TeV $pp$ collider (LHC). The triplet vev is chosen as $v_{\Delta}=10^{-5}$~GeV and the Wilson coefficients are fixed to $C/\Lambda^2=(5~\text{TeV})^{-2}$. We emphasise that the EFT choices at this point are reference values for discussing their phenomenology; we will return to the limits achievable at the LHC below. Note that, for the chosen operators, different values of $v_{\Delta}$ will result in the same cross section, as none of the shown vertices involves the triplet vev in the vertex factors. The blue solid line in Fig.~\ref{fig:Prod_HPPHMM} corresponds to the vanilla type-II contribution mediated via gauge bosons, while the red, green, and cyan solid lines denote the combined contributions from type-II~+~$\mathcal{O}_{G\Delta}$, type-II~+~$\mathcal{O}_{Q \Delta \mathcal{D}}^{(1)}$, and type-II~+~$\mathcal{O}_{u\Delta D}$,~respectively.  

The contribution from the vanilla type-II mechanism is consistently smaller than that arising from the EFT operators for the chosen value of $\Lambda$, as expected. This highlights the ability of non-standard effective interactions to sculpt production cross sections even when $\Lambda$ is pushed to considerably higher scales than the chosen reference value. Among the EFT contributions, $\mathcal{O}_{G\Delta}$ dominates across almost the entire range of considered mass points. This is analogous to similar operators in SMEFT, e.g., influencing SM Higgs pair production via BSM contact interactions~\cite{Dolan:2012ac,LHCHiggsCrossSectionWorkingGroup:2016ypw}. The contribution from $\mathcal{O}_{u\Delta D}$ is slightly smaller than that of $\mathcal{O}_{Q \Delta \mathcal{D}}^{(1)}$, since $\mathcal{O}_{u\Delta D}$ only involves up-type quarks. The difference in the $\mathcal{O}_{G\Delta}$, $\mathcal{O}_{u\Delta \mathcal{D}}$, and $\mathcal{O}_{Q\Delta \mathcal{D}}$ contributions arise due to the difference in the gluon and quark parton distribution functions of the proton.\footnote{Down-type interactions could be considered as well, but for our discussion of inclusive production, this would lead to effective blind directions without gaining a qualitatively novel signature within the boundaries of our analysis. }

%%%%%%%%%%%%%%%%%%%%%%%%%%%%%%%%%%%%%%%%
%%%%%%%%%%%%%%%%%%%%%%%%%%%%%%%%%%%%%%%%

%%%%%%%%%%%%%%%%%%%%%%%%%%%%%%%%%%%%%%%%
%%%%%%%%%%%%%%%%%%%%%%%%%%%%%%%%%%%%%%%%
%%%%%%%%%%%%%%%%%%%%%%%%%%%%%%%%%%%%%%%%
\begin{figure}[b!]
\centering
% ============================================================
% (a) Delta -> l l
% ============================================================
\begin{minipage}[t]{0.19\textwidth}
\centering

\begin{tikzpicture}
\begin{feynman}

    \vertex (aL) at (-1.2,0);
    \vertex (aR) at (0.0,0);

    \draw[scalar] (aL) -- (aR);

    \node at (-0.6,0.3) {$\Delta^{--}$};

    \vertex (b) at (1.6,0.9);
    \vertex (c) at (1.6,-0.9);

    \diagram*{
        (aR) -- [fermion, edge label=$\ell^{-}$] (b),
        (aR) -- [fermion, edge label'=$\ell^{-}$] (c),
    };

\end{feynman}
\end{tikzpicture}

\small{(a)}
\end{minipage}
\hfill
% ============================================================
% (b) Delta -> W W
% ============================================================
\begin{minipage}[t]{0.19\textwidth}
\centering

\begin{tikzpicture}
\begin{feynman}

    \vertex (aL) at (-1.2,0);
    \vertex (aR) at (0.0,0);
    \draw[scalar] (aL) -- (aR);

    \node at (-0.6,0.3) {$\Delta^{--}$};

    \vertex (b) at (1.6,0.9);
    \vertex (c) at (1.6,-0.9);

    \diagram*{
        (b) -- [boson, edge label=$W^{-}$, near start] (aR),
        (c) -- [boson, edge label'=$W^{-}$, near start] (aR),
    };

\end{feynman}
\end{tikzpicture}

\small{(b)}
\end{minipage}
\hfill
% ============================================================
% (c) Delta -> l l Z   (with blob)
% ============================================================
\begin{minipage}[t]{0.19\textwidth}
\centering

\begin{tikzpicture}
\begin{feynman}

    \vertex (aL) at (-0.6,0);
    \vertex (aX) at (0.0,0);
    \vertex[blob, fill=black, minimum size=6pt] (x) at (0.8,0) {};

    \draw[scalar] (aL) -- (aX) -- (x);

    \node at (-0.3,0.3) {$\Delta^{--}$};

    \vertex (b) at (2.0,1.0);
    \vertex (c) at (2.0,-1.0);
    \vertex (d) at (2.0,0);

    \diagram*{
        (x) -- [fermion, edge label=$\ell^{-}$] (b),
        (x) -- [fermion, edge label'=$\ell^{-}$] (c),
        (x) -- [boson, edge label'=$~~~~Z$] (d),
    };

\end{feynman}
\end{tikzpicture}

\small{(c)}
\end{minipage}
\hfill
% ============================================================
% (d) Delta -> l l gamma (with blob)
% ============================================================
\begin{minipage}[t]{0.19\textwidth}
\centering

\begin{tikzpicture}
\begin{feynman}

    \vertex (aL) at (-0.6,0);
    \vertex (aX) at (0.0,0);
    \vertex[blob, fill=black, minimum size=6pt] (x) at (0.8,0) {};

    \draw[scalar] (aL) -- (aX) -- (x);

    \node at (-0.3,0.3) {$\Delta^{--}$};

    \vertex (b) at (2.0,0.9);
    \vertex (c) at (2.0,-0.9);
    \vertex (d) at (2.0,0);

    \diagram*{
        (x) -- [fermion, edge label=$\ell^{-}$] (b),
        (x) -- [fermion, edge label'=$\ell^{-}$] (c),
        (x) -- [photon, edge label'=$~~~~\gamma$] (d),
    };

\end{feynman}
\end{tikzpicture}

\small{(d)}
\end{minipage}
\\[0.4cm]
% ============================================================
% (e) Cascade decay
% ============================================================
\begin{minipage}[t]{0.19\textwidth}
\centering

\begin{tikzpicture}
\begin{feynman}

    \vertex (aL) at (-1.4,0);
    \vertex (a)  at (0.0,0);
    \draw[scalar] (aL) -- (a);

    \node at (-0.7,0.3) {$\Delta^{--}$};

    \vertex (b) at (1.0,1.0);
    \vertex (v) at (1.0,-1.0);

    \vertex (c) at (2.3,-0.3);
    \vertex (d) at (2.3,-1.7);

    \diagram*{
        (a) -- [fermion, edge label=$\ell^{-}$] (b),
        (a) -- [fermion, edge label'=$\ell^{- *}$] (v),
        (v) -- [boson, edge label'=$W^{-}$] (c),
        (v) -- [fermion, edge label'=$\nu_{\ell}$] (d),
    };

\end{feynman}
\end{tikzpicture}

\small{(e)}
\end{minipage}
\hfill
\begin{minipage}[t]{0.69\textwidth}
\vspace{-1cm}
\caption{
Decay topologies of $\Delta^{\pm \pm}$ into leptonic, bosonic, mixed gauge--leptonic, 
and cascade final states.  
Panels (c) and (d) arise from effective operator, represented by black blobs. }
\label{fig:Delta_decay_feyn}
\end{minipage}
\end{figure}
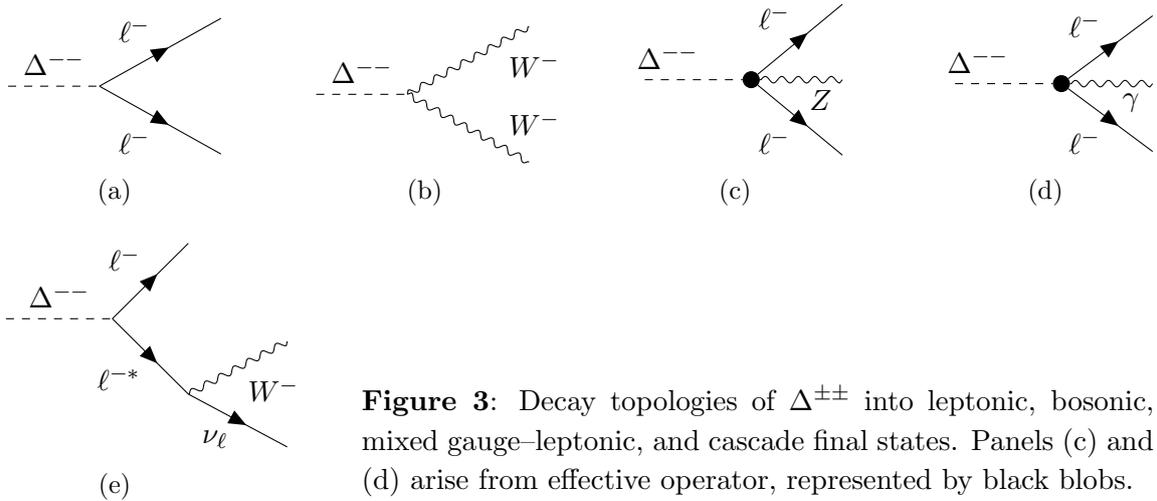
%%%%%%%%%%%%%%%%%%%%%%%%%%%%%%%%%%%%%%%%

The inclusion of the EFT operators not only modifies the production cross section, but can significantly alter the decay of the $\Delta^{\pm \pm}$. As discussed before, $\mathcal{O}_{BL\Delta}$ opens up several other decay modes. It can decay into several two-body and three-body final states, such as, $\ell^{\pm}\ell^{\pm}$, $W^{\pm}W^{\pm}$, $\ell^{\pm}\ell^{\pm}Z$, $\ell^{\pm}\ell^{\pm}\gamma$, and $\ell^{\pm}W^{\pm}\nu_{\ell}$. The Feynman diagrams for these channels are illustrated in Fig.~\ref{fig:Delta_decay_feyn}, shown from left to right in the order listed above. The corresponding branching ratios (BRs) as functions of $M_{\Delta^{\pm \pm}}$ are shown in Fig.~\ref{fig:BR_Delta}. Note that, unlike the production cross section, the decay widths and hence branching ratios of the final states depend on the choice of $v_{\Delta}$. In Fig.~\ref{fig:BR_Delta}, the dashed curves represent the predictions of the vanilla type-II scenario. In contrast, the solid curves correspond to the case where the effective operator $\mathcal{O}_{BL\Delta}$ is included in addition to the vanilla contributions. The choices of  Wilson coefficient $C/\Lambda^2$, and triplet vev are identical as in Fig.~\ref{fig:Prod_HPPHMM}. The red, blue, green, orange and cyan lines denote the BRs of the $\ell^{\pm}\ell^{\pm}$, $W^{\pm}W^{\pm}$, $\ell^{\pm}\ell^{\pm}\gamma$, $\ell^{\pm}W^{\pm}\nu_{\ell}$, and $\ell^{\pm}\ell^{\pm}Z$ channels, respectively. In the vanilla type-II seesaw framework, the same-sign dilepton channel ($\ell^{\pm}\ell^{\pm}$) is governed by the Yukawa interaction. In contrast, the same-sign $W$ boson channel ($W^{\pm}W^{\pm}$) originates from the kinetic term of the doubly-charged Higgs and is proportional to $v_{\Delta}$. The $\ell^{\pm}W^{\pm}\nu_{\ell}$ final state can arise through cascade processes such as $W^{\pm}W^{\pm*} \to \ell^{\pm}W^{\pm}\nu_{\ell}$ or $\ell^{\pm}\ell^{\pm*} \to \ell^{\pm}W^{\pm}\nu_{\ell}$. A third possibility involving $W^{\pm}\Delta^{\pm*}$ is negligible in our case due to the assumed mass degeneracy between $\Delta^{\pm\pm}$ and $\Delta^{\pm}$. For our choice of $v_{\Delta}$, the contribution from off-shell $W^{\pm}$ is suppressed by $v_{\Delta}$, whereas those from off-shell leptons are enhanced via the Yukawa couplings. The $\ell^{\pm}\ell^{\pm}\gamma$ and $\ell^{\pm}\ell^{\pm}Z$ modes emerge  due to the presence of $\mathcal{O}_{BL\Delta}$ operator. 

Since we work in the regime of a small $v_{\Delta}\simeq 10^{-5}$~GeV, the $W^{\pm}W^{\pm}$ channel is highly suppressed compared to the Yukawa-driven modes. This suppression is even more pronounced around $M_{\Delta^{\pm\pm}} \approx 100~\text{GeV}$, where the limited phase space further reduces the branching ratio of the $W^{\pm}W^{\pm}$ mode. Consequently, the $\ell^{\pm}\ell^{\pm}$ channel dominates across the entire mass range for the vanilla type-II seesaw, followed by the photon-dressed $\ell^{\pm}\ell^{\pm}\gamma$ mode. In the $\ell^{\pm}W^{\pm}\nu_{\ell}$ decay, the Yukawa-driven contribution via $\ell^{\pm}\ell^{\pm*}$ is larger than the $v_{\Delta}$-suppressed $W^{\pm}W^{\pm*}$ channel, making it more significant than the $W^{\pm}W^{\pm}$ decay mode.

With the inclusion of the operator $\mathcal{O}_{BL\Delta}$, the $\ell^{\pm}\ell^{\pm}\gamma$ and $\ell^{\pm}\ell^{\pm}Z$ channels receive direct contributions in addition to their standard type-II seesaw origin. For our choice of Wilson coefficient and cut-off scale, the partial decay widths of these channels are significant. The remaining ones remain unaffected and are driven solely by the vanilla type-II interactions. As a result, the $\ell^{\pm}\ell^{\pm}\gamma$ channel quickly becomes the dominant decay mode over nearly the entire $M_{\Delta^{\pm\pm}}$ range. The $\ell^{\pm}\ell^{\pm}Z$ channel appears as the second most dominant mode, though its rate is somewhat suppressed compared to $\ell^{\pm}\ell^{\pm}\gamma$ due to phase-space effects. Around $M_{\Delta^{\pm\pm}} \approx 100,\text{GeV}$, the partial decay width for the $\ell^{\pm}\ell^{\pm} Z$ channel is strongly suppressed due to limited phase space. As $M_{\Delta^{\pm\pm}}$ increases, the phase space opens up and the partial width grows accordingly, which is reflected in the rise of its BR. Although the branching ratios of the decay modes, e.g., $W^{\pm}W^{\pm}$ and $W^{\pm}\ell^{\pm}\nu_{\ell}$, are highly suppressed and therefore not shown in the figure, the hierarchy of decay modes except $\ell^{\pm}\ell^{\pm}\gamma/Z$ follows the same ordering as in the vanilla type-II scenario.

For the $\ell^{\pm}\ell^{\pm}$ decay mode, the branching ratio decreases when increasing $M_{\Delta^{\pm\pm}}$. This occurs as the partial decay width for the $\ell^{\pm}\ell^{\pm} Z$ channel grows with the larger available phase space. From Fig.~\ref{fig:BR_Delta}, it is evident that for triplet mass beyond $200~\text{GeV}$, the branching ratio of $\Delta^{\pm \pm} \to \ell^{\pm} \ell^{\pm}$ falls below $\mathcal{O}(10^{-3})$. For our choice of $C$, $\Lambda$, and $v_{\Delta}$, the $\ell^{\pm}\ell^{\pm}$ decay mode is suppressed across the entire considered $M_{\Delta^{\pm\pm}}$ range once the $\mathcal{O}_{BL\Delta}$ operator is sufficiently large. This is at odds with perturbation theory, but it becomes clear that, for the study of (large) reference values, the phenomenology (and, consequently, experimental analyses) is a priori vulnerable to such sculpting effects. We discuss limits and LHC searches in the next section.

Throughout this work we fix the Wilson coefficient as $C/\Lambda^2=(5~\mathrm{TeV})^{-2}$, corresponding to an EFT cutoff (or new physics) scale $\Lambda=5$ TeV. Although the proton--proton center-of-mass energy at the HL-LHC reaches $\sqrt{s}=14$ TeV, the relevant momentum transfer in the hard scattering process is determined by the partonic center-of-mass energy and is therefore significantly reduced by the parton distribution functions. Following the EFT validity considerations discussed in Refs.~\cite{Busoni:2013lha,Busoni:2014sya}, one may require the characteristic momentum transfer of the process to remain below the cutoff scale up to perturbative factors. For the process $pp\to\Delta^{++}\Delta^{--}$ considered in this work, the momentum transfer can be identified with the invariant mass of the doubly charged scalar pair, namely $Q_{\rm tr}\equiv M_{\Delta^{++}\Delta^{--}}$. We have verified that, for the benchmark points considered, the corresponding invariant-mass distributions are predominantly concentrated in the region satisfying $Q_{\rm tr}/(4\pi)<\Lambda$. This indicates that the dominant contribution to the signal originates from the kinematic regime where the EFT description remains reliable, thereby supporting our choice of $\Lambda=5$ TeV. Nevertheless, as with any effective description, the precise domain of validity ultimately depends on the details of the underlying UV completion.

%%%%%%%%%%%%%%%%%%%%%%%%%%%%%%%%%%%%%%%%
\subsection{Impact on LHC Experimental Constraints}
\label{sec:impl}
\begin{figure}[h]
    \centering
    % First figure
    \begin{subfigure}[b]{0.48\textwidth}
        \centering
        \includegraphics[width=\textwidth]{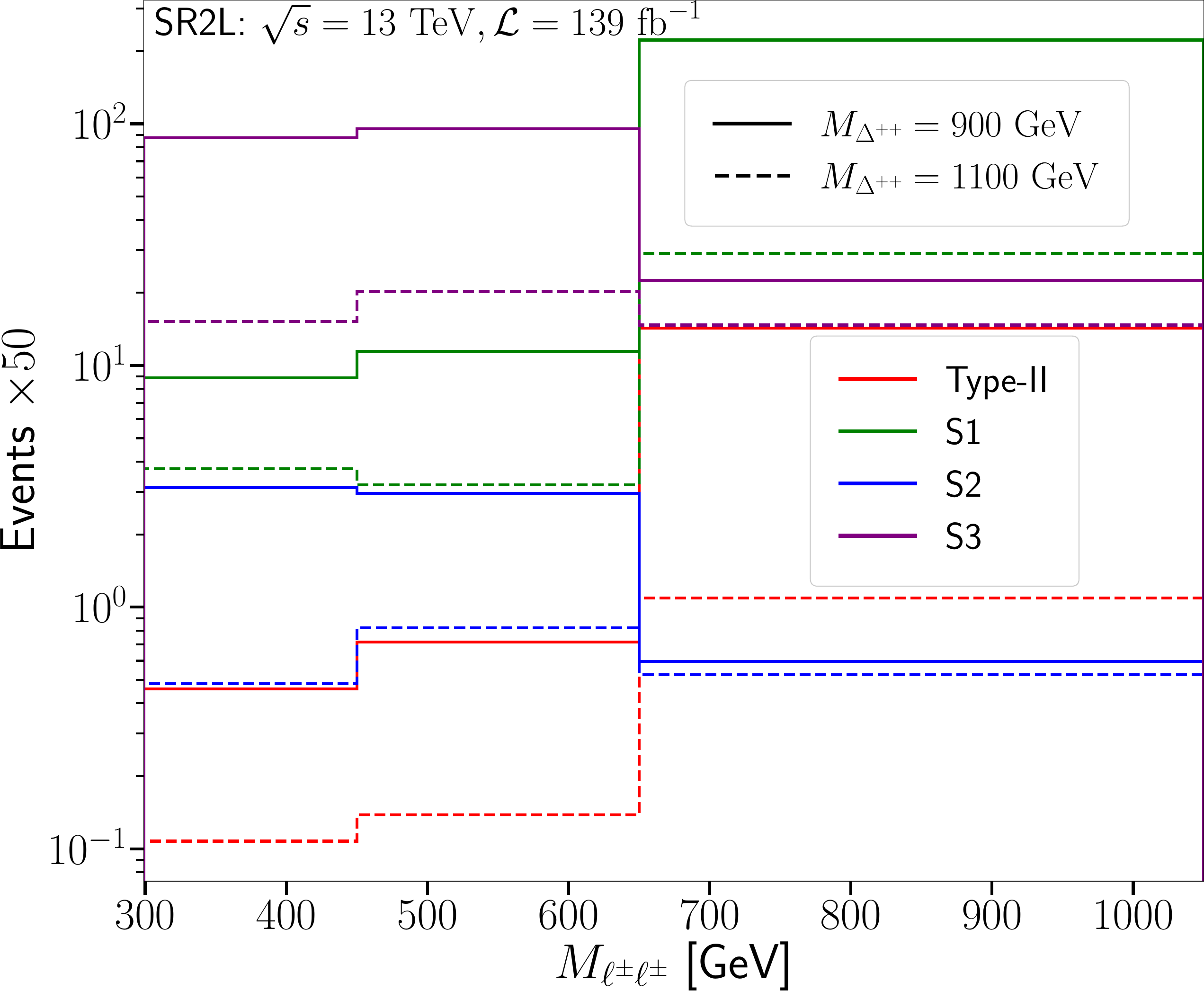}
        \caption{}
        \label{fig:mll_SR2L}
    \end{subfigure}
    \hfill
    % Second figure
    \begin{subfigure}[b]{0.48\textwidth}
        \centering
        \includegraphics[width=\textwidth]{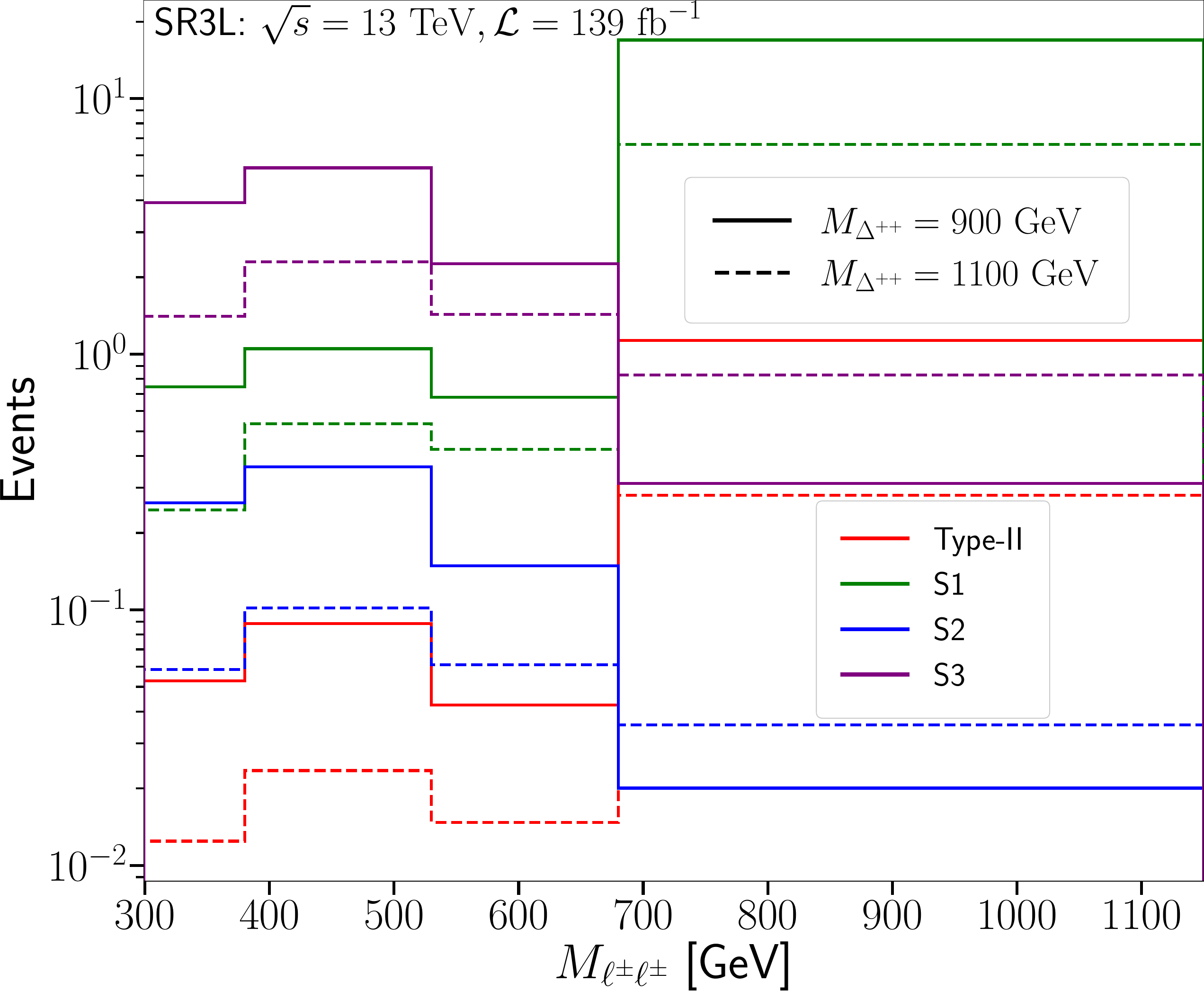}
        \caption{}
        \label{fig:mll_SR3L}
    \end{subfigure}
    \\[0.3cm]
    % Third figure
    \begin{subfigure}[b]{0.48\textwidth}
        \centering
        \includegraphics[width=\textwidth]{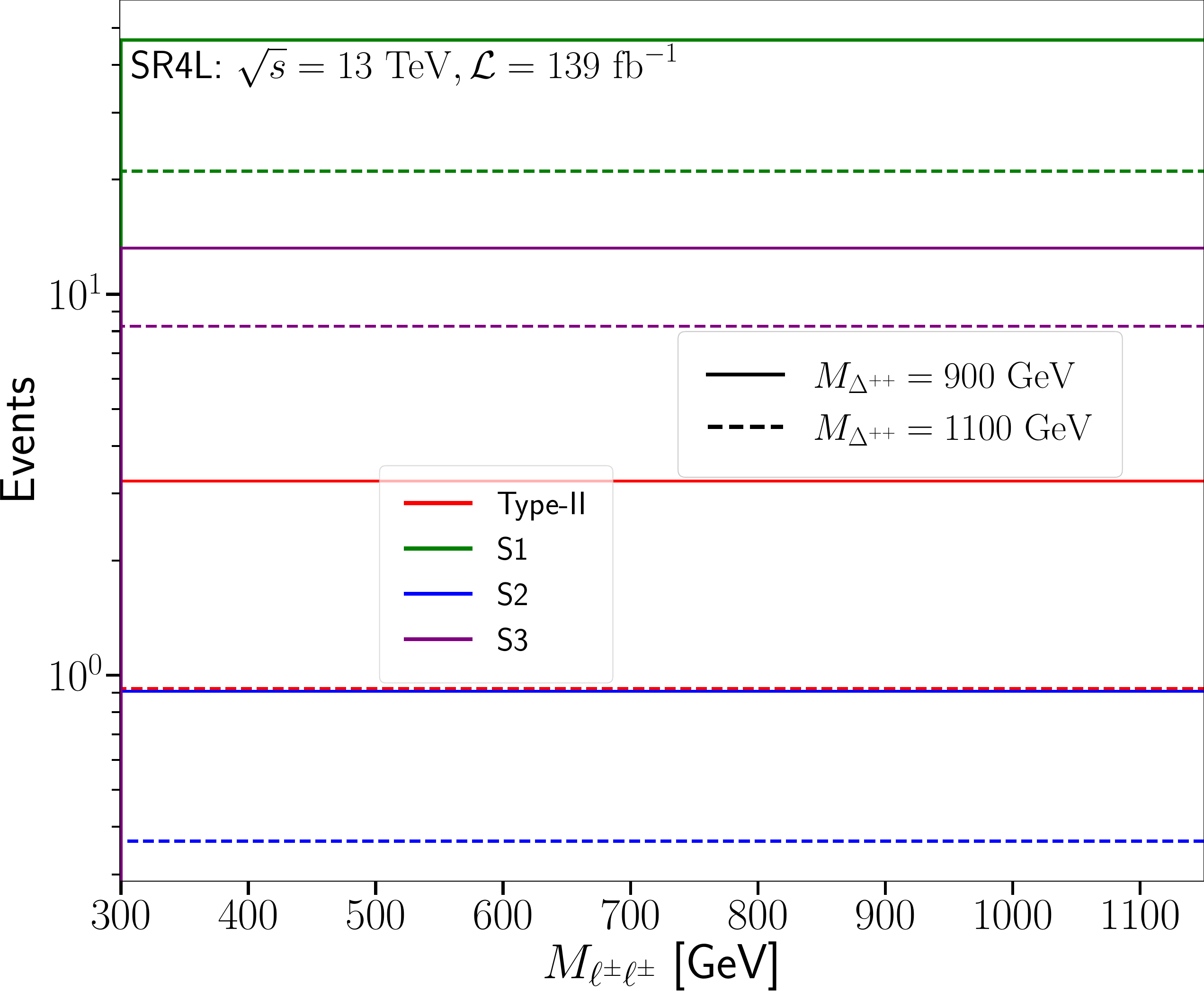}
        \caption{}
        \label{fig:mll_SR4L}
    \end{subfigure}
    \hfill
    \parbox{0.42\textwidth}{\vspace{-6cm}
    \caption{Distributions of the leading same-sign dilepton invariant mass, $m(\ell^{\pm}, \ell^{\prime \pm})_{\text{lead}}$, for the scenarios \textbf{S1}, \textbf{S2}, and \textbf{S3}. The results are shown for two benchmark masses, $M_{\Delta^{\pm\pm}} = 900~\text{GeV}$ and $1100~\text{GeV}$, across the signal regions \textbf{SR2L}, \textbf{SR3L}, and \textbf{SR4L}. }}
    \label{fig:Mll_dis}
\end{figure}

%%%%%%%%%%%%%%%%%%%%%%%%%%%%%%%%%%%%%%%%
The ATLAS collaboration, for instance, has reported on a search for multilepton final states~\cite{ATLAS:2022pbd}, based on an integrated luminosity of $139~\text{fb}^{-1}$ of $pp$ collisions at $\sqrt{s} = 13~\text{TeV}$. This analysis was designed to probe the potential presence of a doubly charged scalar as predicted in the type-II seesaw model. In the remainder of this work, we focus on the two EFT operators introduced in Section~\ref{sec:EFTops}, namely $\mathcal{O}_{G\Delta}$ and $\mathcal{O}_{BL\Delta}$ as phenomenology-driving interactions: The operator $\mathcal{O}_{G\Delta}$ is expected to leave its imprint mostly via the scaling of the pair-production cross section, as the leptons produced from the decay of $\Delta^{\pm \pm}$ easily satisfy the cuts detailed in the ATLAS search. In contrast, the impact of $\mathcal{O}_{BL\Delta}$ is more subtle, as it modifies the kinematic distributions and thereby affects the cut efficiencies, for instance through changes in the key fit variable $m(\ell^{\pm},\ell^{\prime\pm})_{\text{lead}}$~\cite{ATLAS:2022pbd}, defined as the invariant mass of the two leading same-sign leptons. The corresponding distributions are shown in Figs.~\ref{fig:mll_SR2L}, \ref{fig:mll_SR3L}, and \ref{fig:mll_SR4L}. The Wilson coefficients are set to $C/\Lambda^2 = (5~\text{TeV})^{-2}$.

Since the ATLAS search \cite{ATLAS:2022pbd} does not veto any high-$p_T$ photon, the search can still be utilised to derive a limit on the pair-production cross section of doubly-charged scalars, in the presence of a large $\Delta^{\pm \pm} \to \ell^{\pm} \ell^{\pm} \gamma$ mode. This limit can then be readily compared to the theoretical expectation that $\mathcal{O}_{B L \Delta} $ should be a perturbative modification of the vanilla type-II scenario. If the limit is poor, this implies that the current analysis strategy is insufficiently tailored to such an outcome and might miss non-standard type-II versions. 

To this end, we perform a statistical recast of the limit presented in \cite{ATLAS:2022pbd}, accounting for differences in the kinematic distributions relative to the vanilla type-II seesaw and changes in the cut efficiency. Signal events are simulated at leading order (LO) using \texttt{MadGraph5\_aMC@NLO-v3.5.3}~\cite{Alwall:2014hca}. The generated events are passed to \texttt{Pythia8}~\cite{Sjostrand:2006za} for parton showering and hadronisation. For object reconstruction, event selection, and the definition of signal regions, we follow the procedure outlined in Ref.~\cite{ATLAS:2022pbd}. Finally, the statistical interpretation is performed using a hypothesis-testing framework implemented in the \texttt{pyhf} package~\cite{Feickert:2022lzh}, which allows us to derive confidence levels (CLs).

We consider three signal regions, \textbf{SR2L}, \textbf{SR3L}, and \textbf{SR4L}, corresponding to final states with two, three, and four leptons, respectively. The event selection criteria are adopted from Tab.~2 of Ref.~\cite{ATLAS:2022pbd}. In accordance with the same reference, the distribution of $m(\ell^{\pm}, \ell^{\prime \pm})_{\text{lead}}$ is used in the statistical fit to derive the exclusion limits for the following three scenarios:
\begin{cedescription}
\item[S1] Vanilla type-II and operator $\mathcal{O}_{G\Delta}$ is considered, affecting the production of the doubly-charged scalar.  
\item[S2] Vanilla type-II and operator $\mathcal{O}_{BL\Delta}$ is included, contributing to the decay of the doubly-charged scalar into $\ell^{\pm}\ell^{\pm}\gamma$.  
\item[S3] Vanilla type-II and both operators, $\mathcal{O}_{G\Delta}$ and $\mathcal{O}_{BL\Delta}$, are included simultaneously, influencing both the production and decay.  
\end{cedescription}
Figures~\ref{fig:mll_SR2L}, \ref{fig:mll_SR3L}, and \ref{fig:mll_SR4L} display the $m(\ell^{\pm}, \ell^{\prime \pm})_{\text{lead}}$ distributions for \textbf{SR2L}, \textbf{SR3L}, and \textbf{SR4L}, respectively. In each figure, we present the distributions corresponding to the vanilla type-II model and the scenarios S1, S2, and S3, shown in red, green, blue, and violet, respectively. For each scenario, two benchmark masses of the doubly-charged Higgs, $M_{\Delta^{\pm\pm}} = 900$ and $1100~\text{GeV}$, are illustrated using solid and dashed lines, respectively. As is evident from these distributions, the scenarios S1, S2, and S3 exhibit noticeable deviations from the vanilla type-II case. These differences, arising from distinct underlying signal topologies, motivate separate reinterpretations of the exclusion limits for each scenario. For the background, we use the distributions from Ref.~\cite{ATLAS:2022pbd}.
%%%%%%%%%%%%%%%%%%%%%%%%%%%%%%%%%%%%%%
%%%%%%%%%%%%%%%%%%%%%%%%%%%%%%%%%%%%%%

Finally, a statistical combination of all signal regions is performed to set the combined limits.\footnote{Since detector and other systematic effects are not included, a scaling factor is applied by normalising the observed and expected limit obtained in the vanilla type-II scenario to the corresponding ATLAS result, and the same factor is used to rescale the limits for the EFT operator cases.} To construct the combined signal region, we treat all bins across the individual signal regions as statistically uncorrelated. A likelihood model is then built from the observed counts and their associated uncertainties in each bin using the \texttt{uncorrelated\_background} specification in \texttt{pyhf}.
%
%%%%%%%%%%%%%%%%%%%%%%%%%%%%%%%%%%%%%%%%%%%%%%%%%%%%%
\begin{figure}[h]
    \centering
    % First figure
    \begin{subfigure}[b]{0.48\textwidth}
        \centering
       
        \includegraphics[width=\textwidth]{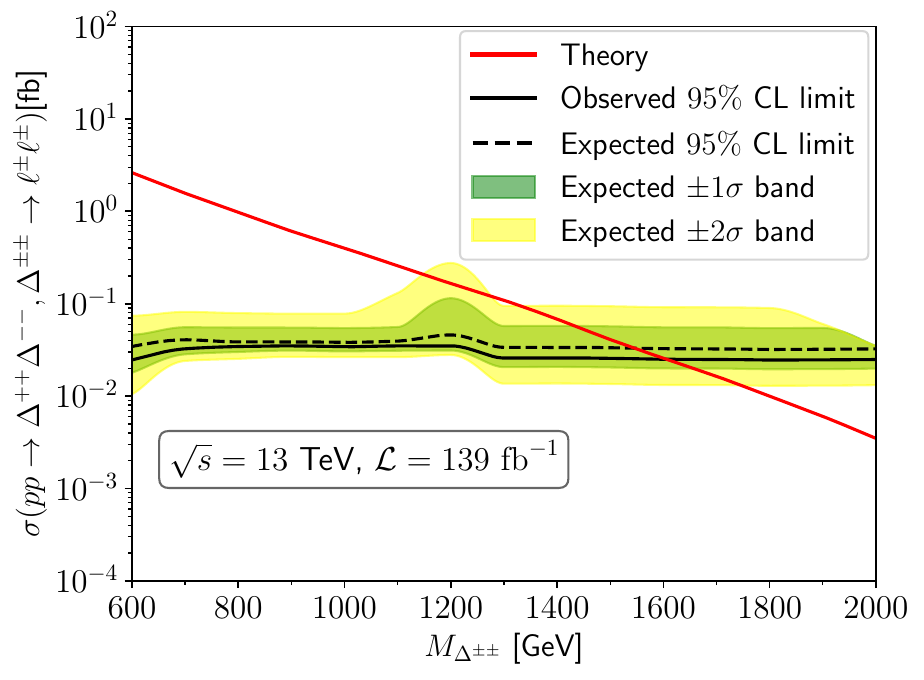}
        \caption{}
        \label{fig:cs1}
    \end{subfigure}
    \hfill
    % Second figure
    \begin{subfigure}[b]{0.48\textwidth}
        \centering
        
        \includegraphics[width=\textwidth]{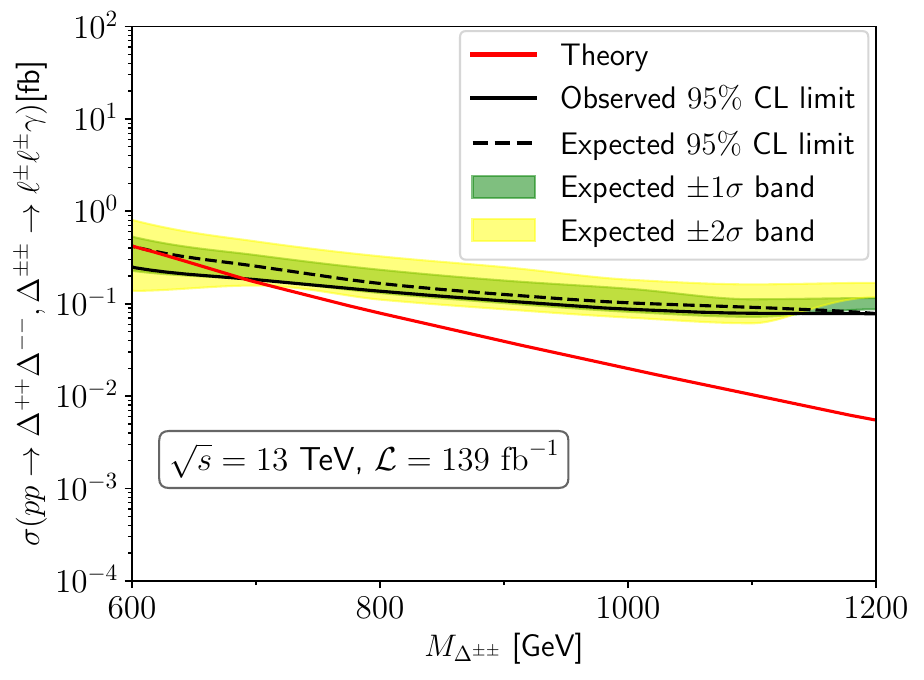}
        \caption{}
        \label{fig:cs2}
    \end{subfigure}
    \\[0.3cm]
    % Third figure
    \begin{subfigure}[b]{0.48\textwidth}
        \centering
       
        \includegraphics[width=\textwidth]{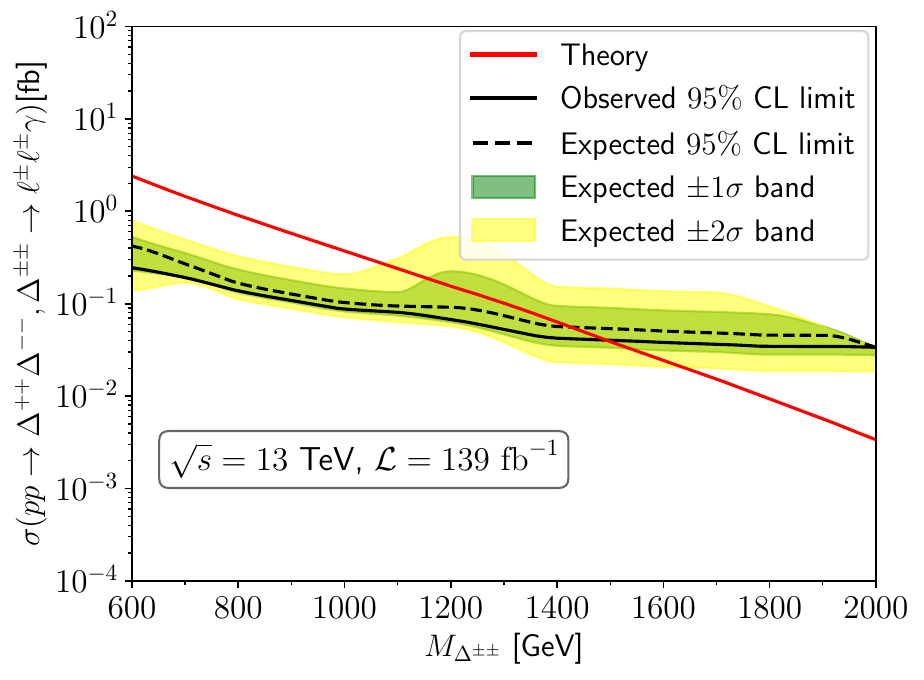}
        \caption{}
        \label{fig:cs3}
    \end{subfigure}
    \hfill
   \parbox{0.42\textwidth}{\vspace{-6cm}
   \caption{Exclusion limits for scenarios \textbf{S1}, \textbf{S2}, and \textbf{S3}, shown in Fig.~\ref{fig:cs1}, Fig.~\ref{fig:cs2}, and Fig.~\ref{fig:cs3}, respectively. The black solid and dashed curves indicate the observed and expected 95\%~CL upper limits, respectively, while the red solid curve shows the leading-order theoretical prediction. The green and yellow bands represent the $\pm1\sigma$ and $\pm2\sigma$ uncertainty intervals around the expected limit.}}
    \label{fig:Exclusion}
\end{figure}
%%%%%%%%%%%%%%%%%%%%%%%%%%%%%%%%%%%%%%%%%%%%%%%%%%%%%
%
The exclusion limit plots corresponding to the three scenarios, \textbf{S1}, \textbf{S2}, and \textbf{S3}, are presented in Figs.~\ref{fig:cs1}-\ref{fig:cs3}, respectively. In each figure, the solid and dashed black curves correspond to the observed and expected limits, respectively, obtained following the procedure described earlier. The red solid curve represents the leading-order (LO) theoretical prediction for the production cross section. The green and yellow bands indicate the $\pm 1\sigma$ and $\pm 2\sigma$ uncertainties on the expected limit, respectively. 

A comparative analysis of the three scenarios reveals distinct exclusion limits on the mass of the doubly-charged scalar. In scenario \textbf{S1}, masses below $M_{\Delta^{\pm \pm}} \lesssim 1600$~GeV are disallowed at 95$\%$ CL. For this parameter choice, we notice that the cuts mentioned in Tab.~2 of Ref.~\cite{ATLAS:2022pbd} are easily satisfied, yielding a large cut-efficiency of $60-70\%$ for the {\bf{SR4L}} signal region. The large cross section, as seen in Fig.~\ref{fig:Prod_HPPHMM}, leads to a very tight constraint on the mass of the doubly-charged Higgs. Scenario \textbf{S2} yields a weaker bound, with masses above $M_{\Delta^{\pm \pm}} \gtrsim 700$~GeV remaining viable. In this case, the production of the doubly-charged scalar is identical to that in the vanilla type-II seesaw. However, the presence of a large $\mathcal{O}_{BL\Delta}$ operator leads to a significant contribution to the $\Delta^{\pm \pm} \to l^{\pm} l^{\pm} \gamma$ decay mode. The different kinematic variables, especially the $m(\ell^{\pm}, \ell^{\prime \pm})_{\text{lead}}$ distribution, is significantly modified as can be seen from Figs.~\ref{fig:mll_SR2L}, ~\ref{fig:mll_SR3L} and ~\ref{fig:mll_SR4L}. This results in a lower cut efficiency and shows a potential vulnerability of current search strategies to modifications of the underpinning dynamics. In contrast, scenario \textbf{S3} which incorporates the enhancement of production cross section due to $\mathcal{O}_{G\Delta}$ operator and difference in $m(\ell^{\pm}, \ell^{\prime \pm})_{\text{lead}}$ distribution due to the presence of $\mathcal{O}_{BL\Delta}$ operator gives rise to a strong constraint, excluding masses below $M_{\Delta^{\pm \pm}} \lesssim 1500$~GeV. The stronger sensitivity to \textbf{S1} and \textbf{S3} again arises from the enhancement of the pair-production cross section of ${\Delta^{\pm \pm}}$, driven by the contribution from the $\mathcal{O}_{G\Delta}$ operator, as illustrated in Fig.~\ref{fig:Prod_HPPHMM}. The bound for \textbf{S2} is much weaker as it takes the different final state like $\ell^{\pm} \ell^{\pm} \gamma $ without any enhancement in production cross section, demonstrating the contamination effect in the $\Delta^{\pm \pm} \rightarrow \ell^{\pm} \ell^{\pm}$ search mentioned above. The slight difference of constraints between \textbf{S1} and \textbf{S3} arises due to the inclusion of the $\mathcal{O}_{BL\Delta}$ operator, which weakens the bounds for \textbf{S3} due to the lower cut efficiency. The impact of the $\mathcal{O}_{G\Delta}$ operator is more pronounced, leading to tighter bounds on $M_{\Delta^{\pm \pm}}$, compared to the relatively weaker effect of the $\mathcal{O}_{BL\Delta}$ operator in shaping the exclusion~limits.

\begin{figure}[!t]
    \centering
    \includegraphics[width=0.62\textwidth]{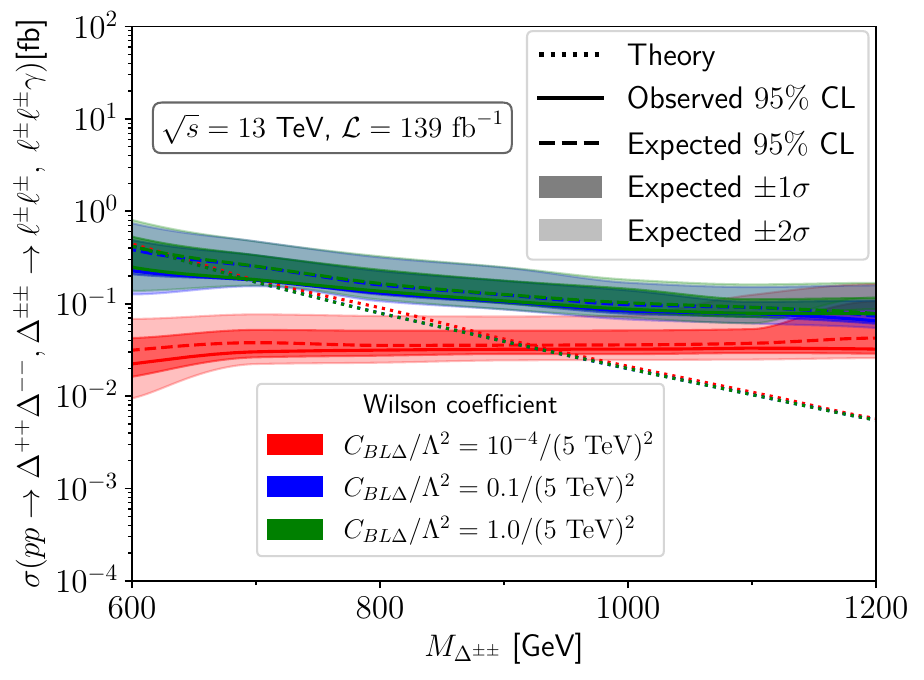}
    \caption{Exclusion limits for scenario \textbf{S2} with variation of $\mathcal{O}_{BL\Delta}$ Wilson coefficient. The red, blue, and green colours correspond to $C_{BL\Delta} = 10^{-4}$, $0.1$, and $1.0$, respectively. The solid (dashed) curves indicate the observed (expected) 95\%~CL upper limits, while the dotted curves show the leading-order theoretical predictions. The dark and light shaded bands represent the $\pm1\sigma$ and $\pm2\sigma$ uncertainty intervals around the expected limit.}
    \label{fig:S2_wilco_vary}
\end{figure}

For the considered scenarios, in particular for \textbf{S2}, we focussed on the extreme choice of the Wilson coefficient of the operator $\mathcal{O}_{BL\Delta}$, namely $C_{BL\Delta}/\Lambda^2=1/(5~\text{TeV})^2$, which corresponds to an almost $100\%$ contribution from the decay mode $\Delta^{\pm\pm} \to \ell^{\pm}\ell^{\pm}\gamma$. To adopt a more general perspective, we now turn to a study of how exclusion limits respond to variation of the Wilson coefficient. More specifically, we include both the $\ell^{\pm}\ell^{\pm}$ and $\ell^{\pm}\ell^{\pm}\gamma$ decay modes of the doubly charged scalar and derive the corresponding exclusion limits for different values of $C_{BL\Delta}$, as shown in Fig.~\ref{fig:S2_wilco_vary}. We observe a reduction in search sensitivity for larger values of the decay Wilson coefficient, $C_{BL\Delta}/\Lambda^2=\{0.1,1.0\}/(5~\text{TeV})^2$. This is expected given our previous discussion: the dilepton invariant mass $m(\ell^{\pm}, \ell^{\prime \pm})_{\text{lead}}$ as the primary fitting variable becomes less effective in this parameter regime due to the presence of additional decay mode $\ell^{\pm} \ell^{\pm} \gamma$. As a result, the derived mass limits are weaker, with an exclusion reach of $M_{\Delta^{\pm\pm}} \simeq 690$~GeV, compared to the small-coupling case $C_{BL\Delta}/\Lambda^2=10^{-4}/(5~\text{TeV})^2$, for which the reduced contamination from the $\ell^{\pm}\ell^{\pm}\gamma$ channel allows for an exclusion up to $M_{\Delta^{\pm\pm}} \simeq 937$~GeV. In the limit of vanishing Wilson coefficient, the result reproduces the stronger bound reported in the experimental analysis, yielding an exclusion of approximately $M_{\Delta^{\pm\pm}} \simeq 1100$~GeV.

%%%%%%%%%%%%%%%%%%%%%%%%%%%%%%%%%%%%%%%%%%%%%%%%%%%%%
%%%%%%%%%%%%%%%%%%%%%%%%%%%%%%%%%%%%%%%%%%%%%%%%%%%%%

If one considers operators such as $\mathcal{O}_{u\Delta D}$ and $\mathcal{O}_{Q \Delta \mathcal{D}}^{(1)}$, they can induce effects similar to those of the $\mathcal{O}_{G\Delta}$ operator by modifying the pair production of $\Delta^{\pm\pm}$. Likewise, the operator $\mathcal{O}_{LeH\Delta D}$ can lead to effects analogous to those arising from $\mathcal{O}_{BL\Delta}$ in the decay of $\Delta^{\pm\pm}$. However, to present a minimal yet representative scenario, we restrict our analysis to the contributions of the $\mathcal{O}_{G\Delta}$ and $\mathcal{O}_{BL\Delta}$ operators throughout this work.

%%%%%%%%%%%%%%%%%%%%%%%%%%%%%%%%%%%%%%%%%%%%%%%%%%%%%
\subsection{Discovery Prospects of a Heavy $\Delta^{\pm \pm}$  at the HL-LHC}
\label{sec:discov}
%%%%%%%%%%%%%%%%%%%%%%%%%%%%%%%%%%%%%%%%%%%%%%%%%%%%%%
Building on our investigations of exclusion, we also analyse the discovery prospects of the doubly-charged scalar for the considered  EFT operators. For simplicity, we ignore the detector effects in this analysis. Firstly, isolated leptons and photons do not pose significant sources of experimental uncertainty and are already under very good control. Secondly, extrapolating to the high-luminosity (HL-)LHC data, we can expect the experimental sensitivity to further improve as more data-driven calibration is applied, directly translating into improved analysis performance. Large improvements with the growing LHC data set have been observed in other searches~\cite{Belvedere:2024wzg}, and adopting a current sensitivity baseline for the HL-LHC appears overly pessimistic to us.

We begin with a cut-based analysis following the \textbf{SR4L} selection criteria outlined in~\cite{ATLAS:2022pbd}, applied to the signal process
\begin{equation}
pp \rightarrow \Delta^{++} \Delta^{--}\text{~with~} \Delta^{\pm\pm} \rightarrow \ell^{\pm} \ell^{\pm} \gamma\,,
\end{equation}
at the HL-LHC. The choice of $v_{\Delta}$, and other parameters $\Lambda$ are identical to those in the previous section. We consider $\mathcal{O}_{G\Delta}$ and $\mathcal{O}_{BL\Delta}$ Wilson coefficient to be $1/(5~\text{TeV})^2$ for illustration purposes. For the backgrounds, we consider the SM processes $VV$, $VV \gamma$, $VV \gamma \gamma$, $VVV$, $VVV \gamma$, $VVV \gamma \gamma$, $t\bar{t}V$, $t\bar{t}V \gamma$, and $t\bar{t}V \gamma \gamma$, where $V = W^{\pm}, Z$. For the $VV$, $VV \gamma$, and $VV \gamma \gamma$ backgrounds, the vector bosons decay exclusively into leptonic final states, while for the remaining backgrounds, all possible decay modes of the intermediate particles are considered. All events were generated in \texttt{MadGraph} with a transverse momentum cut $p_T > 10~\text{GeV}$ applied to the final-state leptons and photons. (These are inclusive compared to the signal selection considered below.)

In Table~\ref{tab:Cutflow}, we present the cut-flow corresponding to the \textbf{SR4L} selection, along with additional cuts on the number of photons to further suppress the background. The mentioned cuts are as follows :
\begin{itemize}
\item \textbf{SR4L} selection: $N_{\text{leptons}} = 4$, $m(\ell^{\pm}, \ell^{\prime \pm})_{\text{lead}} > 300~\text{GeV}$, $\bar{m} > 300~\text{GeV}$ and $M_{\ell^\pm \ell^\mp} \notin [71.2, 111.2]~\text{GeV}$. Here, $N_{\text{leptons}}$ represents the number of leptons, $\bar{m}$ is the average invariant mass of the two same-charge lepton pairs, and $M_{\ell^\pm \ell^\mp}$ is the invariant mass of same-flavour opposite-charge lepton pairs, as defined in the \textbf{SR4L} selection of Tab.~2 of the referenced experimental paper. $m(\ell^{\pm}, \ell^{\prime \pm})_{\text{lead}}$ has been introduced earlier in section~\ref{sec:impl}. The veto $M_{\ell^\pm \ell^\mp} \notin [71.2, 111.2]~\text{GeV}$ significantly reduce the backgrounds involving $Z$ bosons. 
\item $N_{\gamma} \geq 1$: There must be at least one final-state photon with $p_T > 20~\text{GeV}$.

\end{itemize}

%%%%%%%%%%%%%%%%%%%%%%%%%%%%%%%%%%%%%%%%%%%%%%%%%%%%%
\begin{table*}[!t]
	\centering
	\small 
	\resizebox{1\textwidth}{!}{ % <-- resize to 85% width
		\begin{tabular}{||c|c|c|c|c|c|c||}
			\hline 
			& $N_{\text{leptons}} =4 $  & $m(\ell^{\pm}, \ell^{\prime \pm})_{\text{lead}}$ $>$ 300 GeV  & $\bar{m}$ $>$ 300 GeV  & $
M_{\ell^\pm \ell^\mp} \notin [71.2, 111.2]~\text{GeV}$ & $N_{\gamma} \geq 1$  & $\mathcal{Z}$  \\
			\hline
			1500 GeV \, [$7.64 \times 10^{-2}$ fb] & $3.74 \times 10^{-2}$  & $3.12 \times 10^{-2}$ & $2.93 \times 10^{-2}$ & $2.84 \times 10^{-2}$  & $2.84 \times 10^{-2}$ &  $18.98$ \\
			2000 GeV \, [$1.15 \times 10^{-2}$ fb] & $5.95 \times 10^{-3}$  & $5.37 \times 10^{-3}$ & $5.26 \times 10^{-3}$ &$5.13 \times 10^{-3}$  & $5.13 \times 10^{-3}$ &  $5.26 $\\
			\hline \hline
			$VV $  [$6485.21$ fb]     & $1.46$  & $1.93 \times 10^{-1}$ & $1.02 \times 10^{-1}$ & $<6.83\times 10^{-4}  $  & $<6.83\times 10^{-4} $   & \\
			$VV \gamma$  [$34.61$ fb]      & $8.17  \times 10^{-3}$  & $2.08 \times 10^{-3}$ & $9.00 \times 10^{-5}$ & $<6.92\times 10^{-5} $ & $<6.92\times 10^{-5} $  & \\
			$VV \gamma \gamma$  [$1.88 \times 10^{-1}$ fb]  & $3.57 \times 10^{-5}$  & $9.78 \times 10^{-6}$  & $6.39 \times 10^{-6}$ & $<3.76\times 10^{-7} $ & $<3.76\times 10^{-7} $  &  \\
			$VVV $  [$261.36$ fb] &  $7.24 \times 10^{-2}$ & $2.64 \times 10^{-2}$ & $1.72 \times 10^{-2}$& $1.04 \times 10^{-3}$ & $5.23 \times 10^{-4}$  & \\	
			$VVV \gamma$  [$3.04$ fb]  & $1.08 \times 10^{-3}$  & $3.77 \times 10^{-4}$ & $2.19 \times 10^{-4}$ & $9.12 \times 10^{-6}$ & $3.04 \times 10^{-6}$  & \\
            $VVV \gamma \gamma $  [$3.54 \times 10^{-2}$ fb]  & $1.97 \times 10^{-5}$  & $9.02 \times 10^{-6}$ & $5.45 \times 10^{-6}$ & $3.54 \times 10^{-7}$ & $3.54 \times 10^{-7}$ &  \\
            $t \bar{t} V$  [$1112.86$ fb]  & $3.27 \times 10^{-1}$  & $6.54 \times 10^{-2}$ & $2.64 \times 10^{-2}$ & $1.11 \times 10^{-3}$ & $2.78 \times 10^{-4}$  & \\
            $t \bar{t} V \gamma$  [$11.51$ fb]  & $2.85 \times 10^{-3}$  & $5.98 \times 10^{-4}$ & $2.65 \times 10^{-4}$ & $<1.15\times 10^{-5}$ & $<1.15\times 10^{-5}$  & \\
            $t \bar{t} V\gamma \gamma $  [$1.08 \times 10^{-1}$ fb]  & $2.44 \times 10^{-5}$  & $7.92 \times 10^{-6}$ &$3.96 \times 10^{-6}$ & $2.33 \times 10^{-7}$ & $1.16 \times 10^{-7}$  & \\
			\hline 
		\end{tabular}
	}
	\caption{Cross sections for the signal and background processes after imposing the selection cuts, assuming 
$C_{G\Delta}/\Lambda^2 = 1/(5~\mathrm{TeV})^2$ and 
$C_{BL\Delta}/\Lambda^2 = 1/(5~\mathrm{TeV})^2$, 
at an integrated luminosity of 
$L_\text{int.} = 3~\text{ab}^{-1}$ 
corresponding to the HL-LHC. 
    }
	\label{tab:Cutflow}
\end{table*}
%%%%%%%%%%%%%%%%%%%%%%%%%%%%%%%%%%%%%%%%%%%%%%%%%%%%%

As we can see, after imposing \textbf{SR4L} selection and $N_{\gamma} \geq 1$, we obtain more than $5 \sigma $ significance for the two benchmark masses $M_{\Delta^{\pm \pm}} = 1500$ and $2000$ GeV. To evaluate the statistical significance, we use~\cite{Cowan:2010js}\footnote{The significance estimator $\mathcal{Z}$ is based on the asymptotic profile likelihood ratio for a counting experiment and corresponds to the median expected significance evaluated on the Asimov dataset, as derived in Ref.~\cite{Cowan:2010js}. In the asymptotic limit, $\mathcal{Z}$ can be interpreted as the equivalent Gaussian significance associated with the p-value for rejecting the background-only hypothesis.}
\begin{align}
\mathcal{Z} = \sqrt{2\left(N_S+N_B\right)\ln\left(\frac{N_S+N_B}{N_B}\right)-2N_S} \,, 
\end{align}
where $N_S$ and $N_B$ represent the number of signal and background events, respectively. The number of background events is determined using the relation:
\begin{align}
N_B = \left(\sum_{i} \sigma_{B}^{i} \times \epsilon_{B}^{i}\right) \times L_\text{int.}\,,
\end{align}
where $\sigma_{B}^{i}$ is the cross section of the $i^{\text{th}}$ background process, and $\epsilon_{B}^{i}$ denotes its corresponding cut efficiency. The parameter $L_\text{int.} = 3~\text{ab}^{-1}$ is the integrated luminosity of the HL-LHC. The total background yield, $N_B$, is obtained by summing the contributions from all background processes considered~\footnote{After applying the selection cuts described above, no events survive for some of the background processes. This is due to the finite size of the corresponding Monte Carlo samples. To obtain a conservative estimate when evaluating the significance, we assign a background contribution equal to the weight of a single event in such cases. These entries are denoted in the Table~\ref{tab:Cutflow} by the symbol “$<$”, indicating that the quoted value corresponds to an upper bound set by one Monte Carlo event.}.

\begin{figure}[!t]
    \centering
    
    \includegraphics[width=0.62\textwidth]{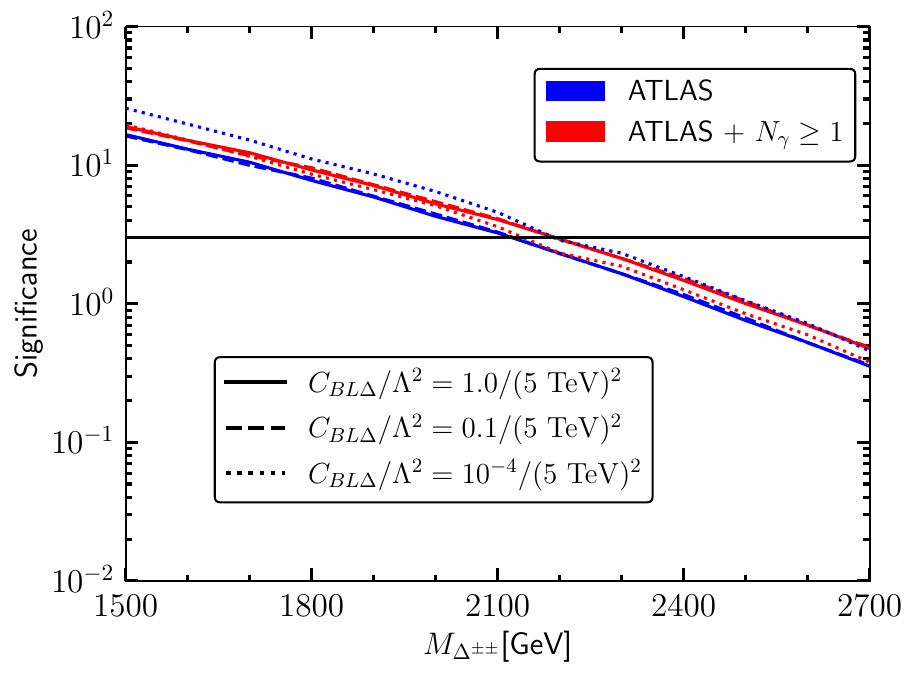}
    \caption{Significance as a function of $M_{\Delta^{\pm\pm}}$ for different sets of cuts mentioned previously in Tab.~\ref{tab:Cutflow} and for different representative $\mathcal{O}_{BL\Delta}$ Wilson coefficient values, evaluated at the 14 TeV HL-LHC with an integrated luminosity of 3000 fb$^{-1}$. }
    \label{fig:Sig_plot_wilco_vary}
\end{figure}

In the previous section, we focused on the $\Delta^{\pm\pm} \to \ell^{\pm}\ell^{\pm}\gamma$ channel, fixing the $\mathcal{O}_{BL\Delta}$ Wilson coefficient to a large value that effectively biased the phenomenology into the three-body decay. We now study the impact of varying this Wilson coefficient and consider the combined contribution from both $\Delta^{\pm\pm} \to \ell^{\pm}\ell^{\pm}$ and $\Delta^{\pm\pm} \to \ell^{\pm}\ell^{\pm}\gamma$ decay modes. The resulting statistical significance as a function of the doubly charged scalar mass $M_{\Delta^{\pm\pm}}$ is shown in Fig.~\ref{fig:Sig_plot_wilco_vary}; the solid, dashed, and dotted lines correspond to Wilson coefficient values of $C_{BL\Delta} = 1.0$, $0.1$, and $10^{-4}$ ($\Lambda=5~\text{TeV}$), respectively. In the plot, the blue and red curves correspond to the significance after applying the \textbf{SR4L} selection and the \textbf{SR4L} selection with $N_{\gamma} \geq 1$, respectively. The solid black line indicates the $3\sigma$ significance threshold. For all three cases, the significance decreases with increasing $M_{\Delta^{\pm\pm}}$. This trend arises because the pair production cross section of $\Delta^{\pm\pm}$ decreases with increasing $M_{\Delta^{\pm\pm}}$, primarily due to phase-space suppression. When only the standard ATLAS selection cuts are applied, the smaller coupling value ($C_{BL\Delta} = 10^{-4}$) yields the highest statistical significance, reflecting the fact that the analysis is optimised for the dilepton decay mode of the charged Higgs. In contrast, when the selection is modified to explicitly require atleast one photons in the final state, larger coupling values ($C_{BL\Delta} = 0.1,\,1.0$) lead to an improved significance. This behaviour is expected, since an increased decay Wilson coefficient enhances the branching fraction into the photonic decay channel. As can be seen, when the \textbf{SR4L} selection along with the $N_{\gamma} \geq 1$ requirement is imposed, the significance reaches the $3\sigma$ level up to $M_{\Delta^{\pm\pm}} = 2200$ GeV for $C_{BL\Delta} = 0.1,\,1.0$. The plot displays the significance starting from $M_{\Delta^{\pm\pm}} = 1500$ GeV, since masses below this value are already constrained by experimental searches~\cite{ATLAS:2022pbd}, as shown in Fig.~\ref{fig:cs3}.

%%%%%%%%%%%%%%%%%%%%%%%%%%%%%%%%%%%%%%%%%%%%%%%%%%%%%
%%%%%%%%%%%%%%%%%%%%%%%%%%%%%%%%%%%%%%%%%%%%%%%%%%%%%

%%%%%%%%%%%%%%%%%%%%%%%%%%%%%%%%%%%%%%%%%%%%%%%%%%%%%
\section{Summary and Conclusions}
\label{sec:conc}
%%%%%%%%%%%%%%%%%%%%%%%%%%%%%%%%%%%%%%%%%%%%%%%%%%%%%
In this work, we have explored an EFT extension of the type-II seesaw framework, where the SM particle content, along with the BSM scalar triplet, is treated as the relevant low-energy degrees of freedom. Among the higher-dimensional operators constructed within this EFT setup, we have focused on those discussed in Sec.~\ref{sec:EFTops}, and have also outlined possible UV completions that could generate them. 

For our phenomenological analysis, we have concentrated on the operators $\mathcal{O}_{G\Delta}$ and $\mathcal{O}_{BL\Delta}$. We reinterpreted the existing ATLAS search for the pair production of $\Delta^{\pm\pm}$ to derive updated limits on the mass $M_{\Delta^{\pm\pm}}$, considering the effects of these operators both individually and simultaneously. Our results demonstrate that including $\mathcal{O}_{G\Delta}$ significantly strengthens the exclusion limit compared to the vanilla type-II seesaw scenario.

Furthermore, we have investigated the discovery prospects at the High-Luminosity LHC (HL-LHC) by examining the newly induced decay channel $\Delta^{\pm\pm} \rightarrow \ell^{\pm} \ell^{\pm} \gamma$. With a simple yet realistic event selection strategy, we find that a statistical significance of up to $3\sigma$ can be achieved for $M_{\Delta^{\pm\pm}}$ values as high as 2.2~TeV. This highlights the potential of EFT-induced operators to enhance both production and decay sensitivities of the doubly-charged Higgs sector in future collider searches.

Although this study has focused on a specific subset of operators within the type-II EFT framework, other higher-dimensional operators may also yield distinct and exotic collider signatures, warranting detailed investigation in future work. The present analysis has been carried out for a few benchmark choices of $v_{\Delta}$ and $\Lambda$, along with the corresponding Wilson coefficients. Varying these parameters could yield qualitatively distinct features and altered signal topologies, offering additional avenues for exploration in forthcoming studies.

%%%%%%%%%%%%%%%%%%%%%%%%%%%%%%%%%%%%%%%%%%%%%%%%%%%%%
\subsection*{Acknowledgements}
%%%%%%%%%%%%%%%%%%%%%%%%%%%%%%%%%%%%%%%%%%%%%%%%%%%%%
We thank Louie Corpe for helpful correspondence and Graeme Crawford for helpful discussions about \texttt{GroupMath}.
C.E. is supported by the Institute for Particle Physics Phenomenology Associateship Scheme. 
M.M. and S.S. gratefully acknowledge the support and hospitality of the University of Glasgow, UK, where this work was initiated. M.M. further acknowledges support from the IPPP DIVA Programme. M.M. and S.S. also acknowledge the use of the \textbf{SAMKHYA} High-Performance Computing Facility at the Institute of Physics (IOP), Bhubaneswar, as well as the two workstations provided by the Institute of Physics, Bhubaneswar through the \textbf{DAE APEX project}, which were extensively used for the numerical computations presented in this work.
W.N. acknowledges support by the Deutsche Forschungsgemeinschaft (DFG, German Research Foundation) under Germany’s Excellence Strategy - EXC 2121 ``Quantum Universe" - 390833306. This work has been partially funded by the Deutsche Forschungsgemeinschaft (DFG, German Research Foundation) - 491245950.

%%%%%%%%%%%%%%%%%%%%%%%%%%%%%%%%%%%%%%%%%%%%%%%%%%%%%
%  Bibliography
%%%%%%%%%%%%%%%%%%%%%%%%%%%%%%%%%%%%%%%%%%%%%%%%%%%%%
\bibliographystyle{JHEP}
\bibliography{references}

\end{document}